\documentclass[sigconf]{acmart}
\AtBeginDocument{%
  }

\usepackage{tabularx}  
\usepackage{amsmath}  

\usepackage{amssymb}  
\usepackage{makecell} 
\usepackage{booktabs} 
\usepackage{threeparttable} 
\usepackage{graphicx} 
\usepackage{multirow} 
\usepackage{longtable}
\usepackage{array} 
\usepackage{wrapfig} 
\usepackage{enumitem}  
\usepackage{hyperref}       
\usepackage{url}            
\usepackage{amsfonts}       
\usepackage{nicefrac}       
\usepackage{microtype}      
\usepackage{xcolor}         

\usepackage{booktabs}

\setcopyright{acmlicensed}
\copyrightyear{2025}
\acmYear{2025}
\acmDOI{XXXXXXX.XXXXXXX}

\acmISBN{978-1-4503-XXXX-X/2018/06}

\begin{document}

\title{A Survey of Datasets and Tasks for Information Diffusion}

\author{Fuxia Guo, Xiaowen Wang, Yanwei Xie, Jingqiu Li, Zehao Wang, Lanjun Wang}
\affiliation{%
  \institution{Tianjin University}
  \city{Tianjin}
  \country{China}
}
\authornote{Corresponding author: wanglanjun@tju.edu.cn.}
\email{wanglanjun@tju.edu.cn}

\renewcommand{\shortauthors}{Guo et al.}

\begin{abstract}
Information diffusion across various new media platforms gradually influences the perceptions, decisions, and social behaviors of individual users.
In information communication studies, the famous Five W’s of Communication model (5W Model) has displayed the process of information diffusion clearly.
At present, although plenty of studies and corresponding datasets have emerged on information diffusion, a systematic categorization of multiple tasks and integration of datasets are still lacking.
To address this gap, we survey a systematic taxonomy of information diffusion tasks and datasets based on the "5W Model" framework.
We first categorize the information diffusion tasks into ten subtasks with definitions, dataset analyses, and representative methodologies, from three main tasks of information diffusion prediction, social bot detection, and misinformation detection.
We also collect the dataset repository of information diffusion tasks with the available links and compare them based on six attributes associated with users and content: user information, social network, bot label, propagation content, diffusion network, and veracity label.
In addition, we discuss the limitations and future directions of current datasets and research topics to advance the future development of information diffusion.
The dataset repository can be accessed on our website \url{https://github.com/fuxiaG/Information-Diffusion-Datasets}.
\end{abstract}





\maketitle

\section{Introduction}
Online information can be extensively diffused across cyberspace as new media enable users to voice opinions, gather public insights, and share engaging content, promoting information diffusion~\cite{Chen19, sabharwal2023survey, bikhchandani1992theory}.
Information diffusion is critical for facilitating online interactions and monitoring social networks~\cite{Xu2022}. Moreover, it can shape individual perceptions, influence public opinion, and potentially lead to economic losses~\cite{ChenSpread21}.

Previous surveys~\cite{infoSurvey2017, Capturing2021, RAZAQUE2022, ZHANG2016,math2023} on information diffusion primarily focus on mechanisms, models, and influencing factors. For instance, \cite{infoSurvey2017} categorizes diffusion models into explanatory and predictive types, highlighting their complementarity, while \cite{Capturing2021} introduces a taxonomy for classifying models across disciplines. Additionally, \cite{RAZAQUE2022} discusses vulnerabilities in diffusion models and their impact on social networks, and \cite{ZHANG2016} provides an overview of theoretical and empirical studies, identifying challenges and future directions. A more recent work~\cite{math2023} offers a comprehensive analysis, combining model categorization with an evaluation of diffusion prediction datasets.
Surveys on specific aspects include \cite{FIRDAUS2018}, which investigates retweet behavior, and \cite{ZhouSurvey2021}, which categorizes methods for predicting information popularity. In the domain of misinformation, \cite{BondielliSurvey2019} reviews techniques for detecting fake news and rumors, while \cite{FARHANGIAN2024} proposes a taxonomy for fake news detection.

Despite these contributions, the current review primarily focus on information diffusion mechanisms from a computer science perspective, failing to consider the practical diffusion situation facing communication studies, diminishing its real-world significance.
Moreover, current studies just discuss single research task. However, many tasks in the information diffusion field are seen as scattered, but actually these issues share similar scenario and are associated with inter-relations. The similar scenario is reflected in the real-world data as a virtual public opinion scenario, but few studies integrate these datasets, not to mention those for multiple tasks.
Therefore, there is still a lack of integrated consideration from the perspective of communication studies in categorizing multiple information diffusion tasks and comparing corresponding datasets in different tasks.

To address this problem, this survey develops a systematic taxonomy for information diffusion tasks utilizing the "5W Model" framework and compiles a set of publicly accessible datasets adapted to each task.
Firstly, the \textit{Five W’s of Communication}\cite{lasswell1948structure}, as a classical communication pattern, elucidates the framework of the information diffusion process. Within this model, we categorize information diffusion tasks into social bot detection (users), misinformation detection (content), and information diffusion prediction (paths, receivers, and effects), aligning with the five "W"s. These primary tasks interact with each other and are further subdivided into ten subtasks as shown in Figure\ref{fg1}.
Secondly, from two entities of user and content in the five "W"s, six common attributes are identified in information diffusion data, presented in Table~\ref{tch2}: user information, bot label, and social network for the user, and propagation content, veracity label, and diffusion network for content. Their presence across renowned datasets is compared in Table~\ref{tDS}.
Thirdly, we elaborate on each renowned subtask shown in Figure~\ref{fg1}, define their research object, review their sixty-five datasets summarized in Table~\ref{tDS}, offer their SOTA methods to know the experimental procedure, and provide the URLs of datasets in Table~\ref{tURL} in Appendix~\ref{chAPP}.
Finally, after collecting and analyzing the datasets, we discuss the limitations and propose future research directions in current datasets and information diffusion tasks.
Therefore, the contributions of our study can be summarized as follows.
\begin{itemize}[leftmargin=0.3cm, itemsep=0cm, topsep=0cm, parsep=0cm, partopsep=0cm]
    \item[$\bullet$]
    We categorize the research tasks in the information diffusion process into three main tasks (social bot detection, misinformation detection and information diffusion prediction) from the three perspectives of the spreader, the content and the combination of paths, receivers and effects proposed in the Five W's of Communication.
    \item[$\bullet$] 
    We further subdivide three main tasks into ten subtasks and compare sixty-five datasets utilizing in these tasks based on six key attributes (user information, social network, bot labels, propagation content, diffusion network, and veracity labels).
    \item[$\bullet$] 
    We identify gaps in current information diffusion datasets and research, proposing directions for improvement. Mainly, comprehensive datasets covering all six attributes are scarce, limiting the simultaneous study of the three main tasks. Existing datasets also lack diversity in language, modality, and source platforms, and many are outdated.    
\end{itemize}

\section{Taxonomy}
\label{ch2}

\subsection{Taxonomy of information diffusion tasks}
\label{2.1}
In 1948, Harold Lasswell, an American political scientist and communication theorist, described an act of communication in~\cite{lasswell1948structure} by answering the following five questions:
\begin{quote}
\textbf{Who}?
Says \textbf{What}?
In \textbf{Which} Channel?
To \textbf{Whom}?
With \textbf{What} Effect?
\end{quote}
This description is known as \textit{Laswell’s Model of Communication} or \textit{Five W’s of Communication} (abbr. 5W Model) due to the five words that begin with the letter $W$ which respectively demonstrate five elements during the process of communication: communicator, message, medium, audience and effect. In other words, it also represents five elements in the diffusion process: user, content, path, receiver, and effect.

\subsubsection{Task description.}

In computer science, research addresses various issues based on the five elements.  
Standing up for the user, it identifies whether the user is normal or a bot~\cite{hayawi2022deeprobot, TwiBot22, weiTwitter19, KUDUGUNTA2018312, qxrBrother23}.  
For the content, it determines whether the information is true or false~\cite{FARHANGIAN2024, ShuFake2017, Improving2023, YANG2024123687, hu2024bad}.  
For the path, it predicts the number of receivers a dissemination path can reach~\cite{Chen19, YangMulti-Scale19, xuCasFlow2023}.  
For the receiver, it identifies the next recipient of the message~\cite{Topological17, yuan2021dyhgcn, DeepInf18, FedInf23}.  
It also explores the influence on the receiver~\cite{QiaoRumor24} and examines the popularity level of the information~\cite{SMP19, abidiPopularity2020, chenMicro2016, Retrieval2024}.  

This paper, following the "5W model" framework, categorizes these issues into three types: social bot detection (users), misinformation detection (content), and information diffusion prediction (paths, receivers, and effects).  
\textit{\textbf{Information diffusion prediction}} (Sec.~\ref{ch3}) predicts future diffusion receivers, paths, and effects, such as user attitudes, the next user in a cascade, and post popularity.  
\textit{\textbf{Social bot detection}} (Sec.~\ref{ch4}) classifies users as bots or humans.  
\textit{\textbf{Misinformation detection}} (Sec.~\ref{ch5}) identifies content as misinformation or real information.  

\subsubsection{Relationship between tasks.}
\begin{figure}
	\centering
	\includegraphics[width=\linewidth]{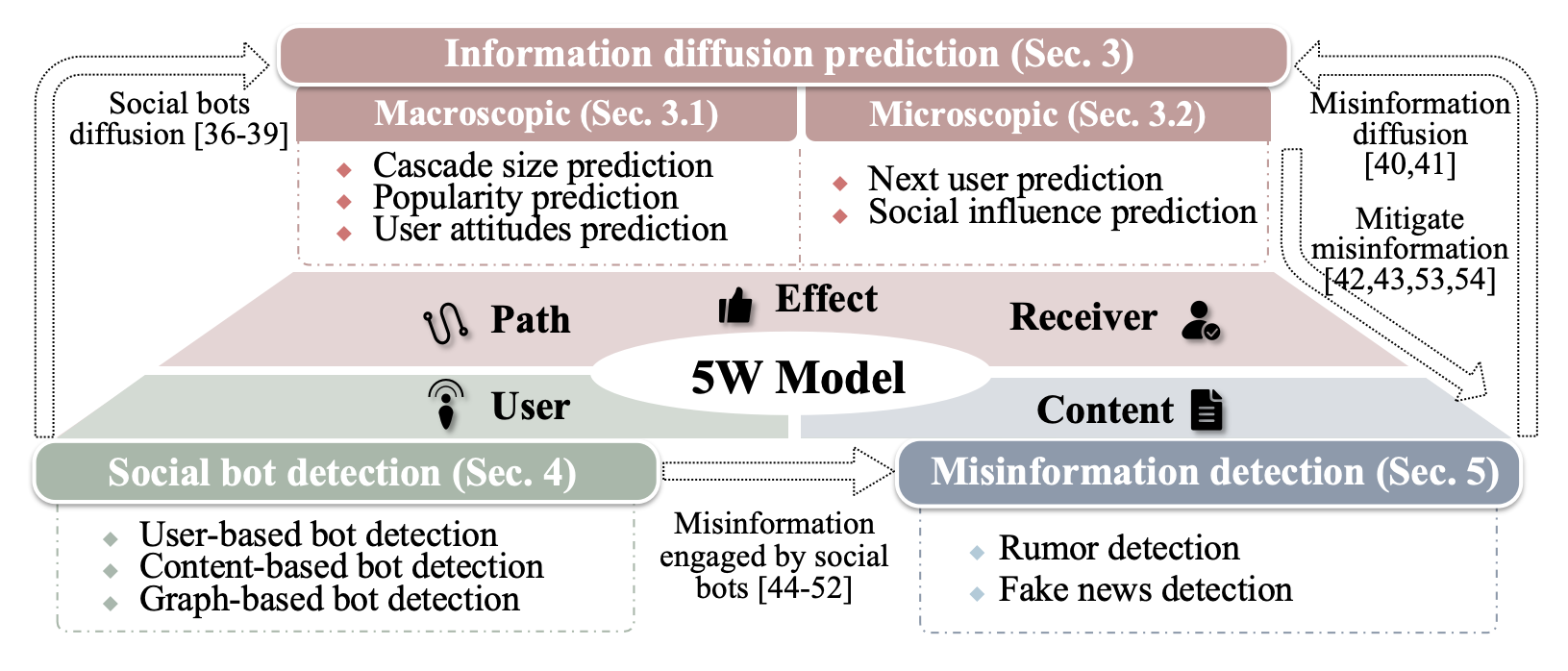}
	\caption{Categorization of information diffusion tasks under the "5W Model" framework includes information diffusion prediction, social bot detection and misinformation detection, along with their respective subtasks and interrelationships.}
	\label{fg1}
\end{figure}

These three main tasks are also interrelated. Some studies~\cite{cai2023network, Ross19, desDetecting22, Evaluating22, king2023diffusion, prandi2020effects, BudakLimiting11, sharma2021network, MinDivide22, dai2020ginger, ShaoSpread17, ShuFakeNewsNet2020, cui2020deterrent, WangFake20, GCAN20, huang2022social, WangAttacking23, Tong18} explore these tasks from overlapping perspectives.
The relationships and cross-studies are illustrated in Figure~\ref{fg1}.
\begin{itemize}[leftmargin=0.3cm, itemsep=0cm, topsep=0cm, parsep=0cm, partopsep=0cm]
    \item[$\bullet$] 
    From social bot detection to information diffusion prediction, there are studies~\cite{cai2023network} analyzing the diffusion ability and characteristics of social bots and exploring information diffusion mechanisms between bots users and human users, and studies~\cite{Ross19, desDetecting22, Evaluating22} assessing the manipulative influence of bots on public opinion.
    \item[$\bullet$]
    From misinformation detection to information diffusion prediction,~\cite{king2023diffusion} analyzes the diffusion characteristics of misinformation, and~\cite{prandi2020effects} models the process of misinformation diffusion with agents.
    \item[$\bullet$]
    From information diffusion prediction to misinformation detection, ~\cite{science2018} accurately predicts the cascade tendency at early stage to assist social platforms in preventing the spread of fake news. Misinformation mitigation and containment research leverage diffusion models such as the independent cascade model~\cite{BudakLimiting11, sharma2021network} and multi-cascade diffusion~\cite{Tong18} which are based on the previous diffusion research on misinformation.
    \item[$\bullet$]
    From social bot detection to misinformation detection, many studies~\cite{MinDivide22, dai2020ginger, ShaoSpread17, ShuFakeNewsNet2020, cui2020deterrent, WangFake20, GCAN20} have shown that social bots contribute to the generation and propagation of misinformation. Further,~\cite{huang2022social} conducts early rumor detection by exploring the behavior of social bots, and~\cite{WangAttacking23} uses adversarial attacks to improve the robustness of fake news detectors by simulating the behavior of bots.
\end{itemize}


\subsection{Taxonomy of information diffusion data attributes}
\label{2.2}

According to the "5W model," information diffusion data involves three entities: user, content, and media. Media, as the data source, encompasses platforms like websites, social media, and video-sharing services, as detailed in Table~\ref{tDS}. 
The attributes necessary for classification stem from two aspects that measure the content within the data. In practice, information diffusion is driven by user engagement, including comments, reposts, clicks, and likes. Unlike traditional media's linear paths, new media operates in a network pattern, where nodes represent users or content and edges denote their relationships. Thus, we can organize user and content attributes based on these nodes and edge relations.
From these perspectives, six key attributes assess information diffusion data: user information, propagation content, bot label, veracity label, social network, and diffusion network. These attributes facilitate dataset comparisons, as shown in the dataset analysis and Table~\ref{tDS}, with corresponding notations in Table~\ref{tch2}.

\subsubsection{Attributes for user.} The attribute information in the user sets involves \textit{user information}, and \textit{bot label} of each user, and static \textit{social network} among users.
\begin{itemize}[leftmargin=0.3cm, itemsep=0cm, topsep=0cm, parsep=0cm, partopsep=0cm]
    \item[$\bullet$] \textit{User information}: Each user $u$ has their user information, including a basic user profile with a combination of semantic information (user description) and property information (numerical and categorical data), and optionally user historical information (posts and user engagements).
    \item[$\bullet$]\textit{Bot label}: There are two types of user accounts: human-operated accounts and AI-controlled social bots accounts, requiring a bot label $y^U$ to denote a social bot user (1) or a human user (0).
    \item[$\bullet$]\textit{Social network}: Among the user nodes set $V^U$, static relationships are denoted as the edges set $E^S$, such as follower/followee relationships, friendships, and co-authorships, which constitute the social network $G^S$.
\end{itemize}

\subsubsection{Attributes for content.} The attribute information in the content sets involves \textit{propagation content} and \textit{veracity label} of each content, and dynamic \textit{diffusion network} among users and its reposts/comments.
\begin{itemize}[leftmargin=0.3cm, itemsep=0cm, topsep=0cm, parsep=0cm, partopsep=0cm]
    \item[$\bullet$]\textit{Propagation content}: Each content $p$ published by users through media, such as posts, news, papers and comments, has its propagation content including texts/images/videos, tags, URLs, etc.
    \item[$\bullet$]\textit{Veracity label}: Propagation content contains both normal information and false misinformation, requiring a veracity label $y^P$ to denote misinformation (1) or normal information (0). In rumor detection, this label sometimes is fine-grained in particular papers.
    \item[$\bullet$]\textit{Diffusion network}: Within a propagation content, dynamic relationships revealing propagation paths are generated by user engagement behaviors among user or content nodes $V=\{V^U, V^P\}$, denoted as the edges set $E^D$, which constitute the diffusion network $G^D$.
\end{itemize}

Diffusion network edges $E^D$ can be further divided into the following three types based on whether nodes represent users or contents in different tasks.
\begin{itemize}[leftmargin=0.3cm, itemsep=0cm, topsep=0cm, parsep=0cm, partopsep=0cm]
    \item[$\bullet$]
    For user-to-user interactions in information diffusion prediction, the edges $E^{D1}$ represent the interaction relationships (click/ like/ repost/ comment/ citation) between two users $u \in V^U$ who publish the propagation content.
    \item[$\bullet$]
    For content-to-content interactions in rumor detection, the edges $E^{D2}$ represent the interaction relationships (repost/ comment/ citation) between a post $p \in V^P$ and its repost/comment or between two reposts/comments.
    \item[$\bullet$]
    For content-to-user interactions in fake news detection, the edges $E^{D3}$ represent the engagement relationships (click/ like/ repost/ comment) between a news item $p \in V^P$ and its engagement elements such as users or between two engagement elements.
\end{itemize}

\begin{table}[h]
  \caption{Notations of the concepts in the datasets}
  \label{tch2}
  \centering
  \begin{tabular}{l|l}  
    \toprule
    \textbf{Symbol}     & \textbf{Description}\\
    \midrule
    $U = \{u_1, u_2, ...\}$ & User set   \\
    $y^U \in \{0, 1\}$    & Bot label \\
    $G^S = \{V^U, E^S\}$    & Social network  \\ 
    $P = \{p_1, p_2, ...\}$ & Content set  \\
    $y^P \in \{0, 1\}$    & Veracity label \\ 
    $G^D = \{V, E^D\}$, $V= \{V^U, V^P\}$ & Diffusion network  \\
    \bottomrule
  \end{tabular}
\end{table}

\setlength{\tabcolsep}{1.5pt}
\begin{table*}
\begin{threeparttable}
    \caption{Datasets comparison for each subtask: user information, social network, bot label, propagation content, diffusion network and veracity label.}
    \label{tDS}
    \centering

    \begin{scriptsize}
    \begin{tabular} {m{0.7cm}|m{0.7cm} | m{3.2cm}|m{3.2cm} | 
    >{\centering\arraybackslash}m{1cm} >{\centering\arraybackslash}m{1cm} >{\centering\arraybackslash}m{1cm} >{\centering\arraybackslash}m{1cm} >{\centering\arraybackslash}m{1cm} >{\centering\arraybackslash}m{1cm}}

    \toprule 
    \multicolumn{2}{c|}{\textbf{Task}} & \textbf{Dataset} & \textbf{Sources} &  \textbf{\makecell[c]{User\\info.}}&  \textbf{\makecell[c]{Social\\network}} &  \textbf{\makecell[c]{Bot\\label}} &  \textbf{\makecell[c]{Prop.\\content}} &  \textbf{\makecell[c]{Diff.\\network}} &  \textbf{\makecell[c]{Veracity\\label}} \\ 
    \midrule

    \multirow{14}{*}{\rotatebox{90}{\makecell[c]{Macroscopic information\\ diffusion prediction}}} &  \multirow{8}{*}{\rotatebox{90}{\makecell[c]{Cascade size\\ prediction}}} & Twitter-casflow~\cite{xuCasFlow2023} & Twitter & — & — & — & — & \checkmark & — \\ 
    ~ & ~ & APS~\cite{xuCasFlow2023} & APS & — & — & — & — & \checkmark & — \\ 
    ~ & ~ & Sina Weibo~\cite{caoDeepHawkes2017} & Sina Weibo & — & — & — & — & \checkmark & — \\ 
    ~ & ~ & HEP-PH~\cite{Graphs05} & arXiv & — & — & — & — & \checkmark & — \\ 
    ~ & ~ & Twitter-FOREST~\cite{YangMulti-Scale19} & Twitter & — & \checkmark & — & — & \checkmark & — \\ 
    ~ & ~ & Douban-FOREST~\cite{YangMulti-Scale19} & Douban & — & \checkmark & — & — & \checkmark & — \\ 
    ~ & ~ & Memetracker~\cite{YangMulti-Scale19} & Google News & — & — & — & — & \checkmark & — \\ \cmidrule(lr){2-10}
    
    ~ & \multirow{5}{*}{\rotatebox{90}{\makecell[c]{Popularity \\prediction}}} & SMPD~\cite{SMP19} & Flicker & \checkmark & — & — & \checkmark & — & — \\ 
    ~ & ~ & Yelp~\cite{linQuantify2022} & Yelp & \checkmark & \checkmark & — & \checkmark & — & — \\ 
    ~ & ~ & MovieLens~\cite{gargOnline2020} & IMDB & \checkmark & — & — & \checkmark & — & — \\ 
    ~ & ~ & Micro-Videos~\cite{chenMicro2016} & Vine & \checkmark & \checkmark & — & \checkmark & — & — \\ 
    ~ & ~ & MicroLens~\cite{niContent2023} & A microvideo platform & — & — & — & \checkmark & \checkmark & — \\ 
    ~ & ~ & ICIP~\cite{ortis2019prediction} & Flicker & \checkmark & — & — & \checkmark & — & — \\ 
    \cmidrule(lr){2-10}
    
    ~ & \multirow{1}{*}{UAP\tnote{1}} & COVID-19-rumor~\cite{chengCOVID192021}  & Twitter, news websites & — & — & — & \checkmark & — & \checkmark \\ \midrule 
    
    \multirow{12}{*}{\rotatebox{90}{\makecell[c]{Microscopic information\\ diffusion prediction}}} & \multirow{6}{*}{\rotatebox{90}{\parbox{1.8cm}{\makecell[c]{Next user \\ prediction}}}} & Twitter-FOREST~\cite{YangMulti-Scale19} & Twitter & — & \checkmark & — & — & \checkmark & — \\ 
    ~ & ~ & Douban-FOREST~\cite{YangMulti-Scale19} & Douban & — & \checkmark & — & — & \checkmark & — \\ 
    ~ & ~ & Memetracker~\cite{YangMulti-Scale19} & Google News & — & — & — & — & \checkmark & — \\ 
    ~ & ~ & Android~\cite{sunMSHGAT2022} & Stack-Exchange & — & \checkmark & — & — & \checkmark & — \\ 
    ~ & ~ & Christianity~\cite{sunMSHGAT2022} & Stack-Exchange & — & \checkmark & — & — & \checkmark & — \\ 
    ~ & ~ & Twitter-MSHGAT~\cite{sunMSHGAT2022} & Twitter & — & \checkmark & — & — & \checkmark & — \\ 
    ~ & ~ & Douban-MSHGAT~\cite{sunMSHGAT2022} & Douban & — & \checkmark & — & — & \checkmark & — \\ 
    ~ & ~ & Douban-ComSoc~\cite{ComSoc2012} & Douban & — & \checkmark & — & — & \checkmark & — \\ 
    \cmidrule(lr){2-10} 
    
    ~ &  \multirow{6}{*}{\rotatebox{90}{\parbox{1.3cm}{\centering \makecell[c]{Social influence\\ prediction}}}} & OAG-DeepInf~\cite{DeepInf18} & MAG, AMiner & — & \checkmark & — & \checkmark & \checkmark & — \\ 
    ~ & ~ & Digg-DeepInf~\cite{DeepInf18} & Digg & — & \checkmark & — & — & \checkmark & — \\ 
    ~ & ~ & Twitter-DeepInf~\cite{DeepInf18} & Twitter & — & \checkmark & — & — & \checkmark & — \\ 
    ~ & ~ & Higgs Twitter~\cite{de2013anatomy} & Twitter & — & \checkmark & — & — & \checkmark & — \\ 
    ~ & ~ & Weibo-DeepInf~\cite{DeepInf18} & Sina Weibo & — & \checkmark & — & — & \checkmark & — \\ 
    ~ & ~ & Weibo-influencelocality~\cite{ZhangInfluence13} & Sina Weibo & \checkmark & \checkmark & — & \checkmark & \checkmark & — \\ \midrule 
    
     \multirow{8}{*}{\rotatebox{90}{\parbox{2.5cm}{\centering Social bot detection}}} &  \multirow{3}{*}{\rotatebox{90}{\parbox{0.8cm}{\centering \makecell[c]{User-\\based }}}} & Cresci-2017~\cite{Cresci17} & Twitter & \checkmark & — & \checkmark & \checkmark & — & — \\ 
    ~ & ~ & gilani-2017~\cite{Gilani17} & Twitter & \checkmark & — & \checkmark & — & — & — \\ 
    ~ & ~ & botometer-feedback-2019~\cite{yangArming2019} & Twitter & \checkmark & — & \checkmark & — & — & — \\ \cmidrule(lr){2-10} 
    
    ~ &  \multirow{2}{*}{\rotatebox{90}{\parbox{0.7cm}{\centering \makecell[c]{Content-\\based }}}} & PAN-AP-2019~\cite{rangel2019overview} & Twitter & — & — & \checkmark & \checkmark & — & — \\[0.1cm] 
    ~ & ~ & caverlee-2011~\cite{leeSeven2021} & Twitter & \checkmark & — & \checkmark & \checkmark & — & — \\[0.1cm]  \cmidrule(lr){2-10} 
    
    ~ &  \multirow{3}{*}{\rotatebox{90}{\makecell[c]{Graph-\\based }}} & Cresci-2015~\cite{CRESCI2015} & Twitter & \checkmark & \checkmark & \checkmark & — & — & — \\ 
    ~ & ~ & TwiBot-20~\cite{fengTwiBot20} & Twitter & \checkmark & \checkmark & \checkmark & — & — & — \\ 
    ~ & ~ & TwiBot-22~\cite{TwiBot22} & Twitter & \checkmark & \checkmark & \checkmark & — & — & — \\ \midrule 
    
     \multirow{23}{*}{\rotatebox{90}{Misinformation detection}} &  \multirow{9}{*}{\rotatebox{90}{Rumor detection}} & PHEME-v1~\cite{zubiagaLearning2016} & Twitter & \checkmark & — & — & \checkmark & \checkmark & \checkmark  \\ 
    ~ & ~ & PHEME-v2~\cite{kochkinaAll2018} & Twitter & \checkmark & — & — & \checkmark & \checkmark & \checkmark  \\
    ~ & ~ & PHEME-v3~\cite{ZHENG2023} & Twitter & \checkmark & \checkmark & — & \checkmark & \checkmark & \checkmark  \\ 
    ~ & ~ & Weibo-BiGCN~\cite{bian2020rumor} & Sina Weibo & — & — & — & — & \checkmark & \checkmark   \\ 
    ~ & ~ & Ma-Weibo~\cite{ma2016detecting} & Sina Weibo & \checkmark & — & — & \checkmark & \checkmark & \checkmark   \\ 
    ~ & ~ & Twitter15~\cite{maDetect2017} & Twitter & — & — & — & \checkmark & \checkmark & \checkmark \\ 
    ~ & ~ & Twitter16~\cite{maDetect2017} & Twitter & — & — & — & \checkmark & \checkmark & \checkmark  \\ 
    ~ & ~ & Twitter15-RDMSC~\cite{ZHENG2023} & Twitter & \checkmark & \checkmark & — & \checkmark & \checkmark & \checkmark   \\ 
    ~ & ~ & Twitter16-RDMSC~\cite{ZHENG2023} & Twitter & \checkmark & \checkmark & — & \checkmark & \checkmark & \checkmark  \\ 
    ~ & ~ & MR2~\cite{HuMR223} & Social media platforms & — & — & — & \checkmark & — & \checkmark \\ \cmidrule(lr){2-10} 
    
    ~ &  \multirow{14}{*}{\rotatebox{90}{Fake news detection}} & FakeNewsNet~\cite{ShuFakeNewsNet2020} & Politifact, Gossipcop & \checkmark & \checkmark & — & \checkmark & \checkmark & \checkmark \\ 
    ~ & ~ & FakeNewsNet-DECOR~\cite{WuDECOR23} & Politifact, Gossipcop & \checkmark & — & — & — & — & \checkmark \\ 
    ~ & ~ & FakeNewsNet-UPFD~\cite{DouPreference21} & Politifact, Gossipcop & \checkmark & — & — & \checkmark & \checkmark & \checkmark \\ 
    ~ & ~ & TruthSeeker2023~\cite{Dadkhah23} & PolitiFact & \checkmark & — & \checkmark & — & — & \checkmark \\ 
    ~ & ~ & MC-Fake~\cite{MinDivide22} & Twitter & \checkmark & \checkmark & — & — & \checkmark & \checkmark \\ 
    ~ & ~ & FineFake~\cite{zhou2024finefake} & Snopes & \checkmark & — & — & \checkmark & \checkmark & \checkmark \\
    ~ & ~ & FauxBuster~\cite{FauxBuster18} & Twitter, Reddit & — & — & — & \checkmark & \checkmark & \checkmark \\
    ~ & ~ & MM-Covid~\cite{li2020mmcovid} & PolitiFact, FullFact & \checkmark & — & — & \checkmark & \checkmark & \checkmark \\
    ~ & ~ & MuMIN~\cite{MuMiN22} & Twitter & — & — & — & \checkmark & \checkmark & \checkmark \\
    ~ & ~ & CHECKED~\cite{yang2021checked} & Sina Weibo & — & — & — & \checkmark & — & \checkmark \\ 
    ~ & ~ & FakeSV~\cite{qiFakeSV2023} & Douyin, Kuaishou & \checkmark & — & — & \checkmark & — & \checkmark \\ 
    ~ & ~ & FTT~\cite{huLearn2023} & Fake news detection system & — & — & — & \checkmark & — & \checkmark \\ 
    ~ & ~ & MCFEND~\cite{liMCFEND2024} & Fact-checking agencies & \checkmark & — & — & \checkmark & — & \checkmark \\ 
    ~ & ~ & Weibo21~\cite{Weibo21} & Sina Weibo & — & — & — & \checkmark & — & \checkmark \\ 
    ~ & ~ & Image-verification-corpus~\cite{boididou2018detection} & Twitter & \checkmark & — & — & \checkmark & — & \checkmark \\ 
    ~ & ~ & Breaking!~\cite{2019breaking} & BS Detector & — & — & — & \checkmark & — & \checkmark \\ 
    ~ & ~ & LIAR~\cite{LiarWang17} & Politifact & — & — & — & \checkmark & — & \checkmark \\ 
    ~ & ~ & Evons~\cite{2022evons} & Media-source & — & — & — & \checkmark & — & \checkmark \\ 
    ~ & ~ & WeChat~\cite{WeakWang2020} & WeChat’s Official Accounts & — & — & — & \checkmark & — & \checkmark \\ 
    ~ & ~ & Fakeddit~\cite{nakamura2020fakeddit} & Reddit & — & — & — & \checkmark & — & \checkmark \\ 

    \bottomrule
    \end{tabular}

    \begin{tablenotes}
    \item[1] UAP: User attitudes prediction
    \end{tablenotes}
\end{scriptsize}
\end{threeparttable}
\end{table*}

\section{Information diffusion prediction}
\label{ch3}
The prediction of information diffusion is used to forecast future propagation paths, users, and effects. 
In terms of output scale, information diffusion prediction can be focused on two directions~\cite{YangMulti-Scale19}: macroscopic information diffusion prediction (Sec.~\ref{3.1}), which predicts the general situation of a social group or the entire environment, and microscopic information diffusion prediction (Sec.~\ref{3.2}), which predicts the behavior of individual users during the diffusion process. We will elaborate on these directions through specific subtasks, providing their definitions, associated datasets, and representative methods.


\subsection{Macroscopic information diffusion prediction}
\label{3.1}
From a macroscopic view, propagation content diffuses widely over time, leading users to form different perceptions based on the information received. For the contents, current prediction problems focus on anticipating graph-based cascade size~\cite{Chen19, YangMulti-Scale19, xuCasFlow2023} and content-based popularity~\cite{SMP19, abidiPopularity2020, chenMicro2016}. For user groups, the prediction problem involves forecasting user attitudes~\cite{QiaoRumor24}. Thus, there are three subtasks in macroscopic information diffusion prediction: cascade size prediction, popularity prediction, and user attitudes prediction.

Greater user engagement with a post indicates higher attention. Predicting the future number of participants (cascade size) reflects the level of concern for the post. \textbf{Cascade size prediction} estimates the scale of information cascades by forecasting the total number of users engaged in the cascade~\cite{YangMulti-Scale19}.
Some popular content receives extensive views and is widely spread, while much content garners little attention. Besides user numbers, the content itself can indicate a post's popularity. Therefore, \textbf{popularity prediction} forecasts online posts' popularity scores as a regression problem by analyzing multimodal propagation content (text, image, and video), spatio-temporal data, and user information~\cite{SMP19}. 
Additionally, harmful content like can evoke negative emotions in users, leading to social and economic chaos~\cite{ChenSpread21}. To understand these effects, \textbf{user attitudes prediction} assesses crowd attitudes during rumor propagation by modeling rumor dynamics~\cite{QiaoRumor24}.

\subsubsection{Definition.}
Let $U$ be the set of users and $P$ be the set of posts.
User engagement behavior can generate a \textit{cascade sequence} $C(t_w)=\{(u_{i}, u_{j}, t_{j}) \}_{i,j \le L}$, which means that there are $L$ users interacting with the initial post $p$ by the observed time $t_w$. The triple $(u_{i}, u_{j}, t_{j})$ indicates that user $u_{j}$ interacts with user $u_{i}$'s post at timestamp $t_{j}$.
From the interaction relations between two users in $C(t_w)$, we can construct a \textit{diffusion network} $G^{D}(t_w)=\{V^{U}, E^{D1}\}$ at time $t_w$.
In particular, for user attitude prediction during rumor propagation, \textit{user crowds} can be categorized into three groups based on classical epidemic models SI, SIS, and SIR~\cite{RevModPhys09}: ignorants ($X$) who are unaware of the rumor, spreaders ($Y$) who spread it, and stiflers ($Z$) who know it but do not spread it. Stiflers can be further divided into neutrality ($Z0$), belief ($Z1$), and disbelief ($Z2$)~\cite{QiaoRumor24}. Predictions using epidemic models require the \textit{group density} of ignorants ($x(t)$), spreaders ($y(t)$), and the three types of stiflers ($z_0(t)$, $z_1(t)$, and $z_2(t)$) at time $t$.
Thus, based on the symbol definitions above, we can define cascade size prediction, popularity prediction, and user attitude prediction as follows.
\begin{enumerate}[label=(\arabic*), leftmargin=0.3cm, itemsep=0cm, topsep=0cm, parsep=0cm, partopsep=0cm]
  \item[$\bullet$]\textit{Cascade size prediction}: Given the cascade sequence $C(t_w)$ and the diffusion network $G^{D}(t_w)=\{V^{U}, E^{D1}\}$ of an initial post $p$ at observed timestamp $t_w$, cascade size prediction aims to predict the size of the cascade $|C(t_p)|$ at predicted timestamp $t_p \gg t_w$.
  \item[$\bullet$]\textit{Popularity prediction}: Given the post set $P$ with post content and the user set $U$, popularity prediction aims to predict the popularity $\omega^{i}$ of a post $p^{i}\in P$.
  \item[$\bullet$]\textit{User attitudes prediction}: Given a rumor $p$, the density of its ignorants $x(t_0)$ and the density of its spreaders $y(t_0)$ at the initial timestamp $t_0$, user attitudes prediction aims to obtain the group density $z_i(t_p)$ of stiflers $z_i$ at predicted timestamp $t_p$.
\end{enumerate}

\subsubsection{Datasets analysis.} The datasets of macroscopic information diffusion prediction are compared in Table~\ref{tDS}.
\begin{itemize}[leftmargin=0.3cm, itemsep=0cm, topsep=0cm, parsep=0cm, partopsep=0cm]
    \item[$\bullet$] 
    The source platforms include social media platforms primarily, academic citation websites in cascade size prediction, and review sites of movies or merchants in popularity prediction. Social media is the most common platform in information diffusion. Academic citation datasets, focused on citation patterns, highlight how information propagates in scholarly contexts, where influence and relevance drive cascade sizes. Review sites reflect user opinions and behavior trends, which are critical for understanding popularity but might lack the intricate interaction details found in social media.
    \item[$\bullet$] 
    The datasets in cascade size prediction all include the diffusion network, while Twitter-FOREST and Douban datasets include social network additionally. It helps capture the influence of user relationships during propagation, potentially supplementing the underlying user patterns not in diffusion networks.
    \item[$\bullet$] 
    The multimodality in SMPD, Yelp and ICLP datasets includes texts and images. Micro-Videos includes texts and videos. MicroLens incorporates texts, images, and videos. This variety improves predictive performance by leveraging multiple types of information.
\end{itemize}

\subsubsection{Methodology.} To demonstrate the method of macroscopic information diffusion prediction and the current performance measured by evaluation metrics, we will present a SOTA method for each subtask that introduces its landmark algorithm, respectively. Their performance are shown in Appendix~\ref{chPerf}.
\begin{itemize}[leftmargin=0.3cm, itemsep=0cm, topsep=0cm, parsep=0cm, partopsep=0cm]
    \item[$\bullet$] \textit{Cascade size prediction}: 
    \textbf{CasFlow}\cite{xuCasFlow2023} proposes a probabilistic cascade size prediction model integrating hierarchical structures and propagation uncertainty. The model first analyzes both local and global diffusion patterns, then captures user interactions over time with Bi-GRUs. Following this, the model encodes uncertainty in propagation using Variational Autoencoders (VAEs) and refines predictions with Normalizing Flows.
    Table~\ref{tb3.1.1} presents its performance comparison between CasFlow and baseline models across three datasets (Twitter-casflow~\cite{xuCasFlow2023}, APS~\cite{xuCasFlow2023}, and Sina Weibo~\cite{caoDeepHawkes2017}) with different observation times, measured by MSLE and MAPE.
    \item[$\bullet$] \textit{Popularity prediction}: 
    \textbf{RAGTrans}\cite{Retrieval2024} introduces a retrieval-augmented model for predicting the popularity of multimodal social media content. It retrieves relevant instances from a user-generated content (UGC) memory bank, builds a multimodal hypergraph, and applies a bootstrapping transformer for neighborhood aggregation. After that, a user-aware fusion module combines multimodal data with user characteristics. 
    Table~\ref{tb3.1.2} presents its performance comparison between RAGTrans and baseline models on three datasets (SMPD~\cite{SMP19}, ICIP~\cite{ortis2019prediction}, and WeChat~\cite{Retrieval2024}), measured by MSE, MAE and SRC.
     \item[$\bullet$]
    \textit{User attitudes prediction}:
    \textbf{Neutral state model}\cite{QiaoRumor24} introduces a \textbf{neutral state model} to represent the crowd attitudes during rumor propagation by segmenting individuals into Ignorants, Spreaders, Skeptics, and Stifflers with varying beliefs. The model uses dynamic equations to simulate the flow of individuals and incorporates parameters for rumor spread rate and the influence of neutral discussions.
    Table~\ref{tb3.1.3} shows its simulated results of neutral state model and XYWZ1Z2 model compared with the actual data on the COVID-19-rumor~\cite{chengCOVID192021} dataset measured by MAE and MSE.
\end{itemize}

\subsection{Microscopic information diffusion prediction}
\label{3.2}
Microscopic information diffusion prediction targets the future diffusion result that a certain user will engage with certain content at a predicted time.
From the standpoint of this content, the study of which user will engage with the target content is known as next user prediction~\cite{YangMulti-Scale19, Topological17, yuan2021dyhgcn}.
From the standpoint of this user, the study of whether the content will be engaged by the target user is known as social influence prediction~\cite{DeepInf18, FedInf23}.
Therefore, there are two subtasks of macroscopic information diffusion prediction, involving next user prediction and social influence prediction.

Intuitively, a user is more likely to repost or comment on content from users they follow or who follow them, especially if they share the same dynamic diffusion trace~\cite{yuan2021dyhgcn}. These relationships and historical diffusion records are available in social and diffusion networks. Thus, based on these networks, \textbf{next user prediction} forecasts which user is likely to be the next to engage in a cascade~\cite{YangMulti-Scale19}. 
From a different perspective of a user, we wonder whether the content will be engaged by this user. \textit{Social action} in activities are  citations in academic sites, voting in news platforms, and reposting or commenting on social media~\cite{FedInf23}. Research~\cite{DeepInf18} suggests that users' emotions, decisions, and actions are influenced primarily by their social network neighbors, without external disturbances. Therefore, \textbf{social influence prediction} is necessary to forecast changes in a user's actions regarding certain content based on their neighbors' influence. So similarly, social influence prediction also needs the social network and diffusion network.

\subsubsection{Definition.}
In the social influence prediction task, \textit{$\rho$-neighbors} $\Gamma^\rho_{u_i} = \{ u_j | d(u_j, u_i) \le \rho, \, j \neq i  \}$ of user $u_i \in U$ are required to represent users whose shortest path distances from $u_i$ in the social network $G^{S}=\{V^{U}, E^{S}\}$ are no more than $\rho$. User $u_i$ is referred to as the ego user of these neighbors.
$A^t_{u_i} = \{a^{t}_{u_j} | u_j \in \Gamma^\rho_{u_i}\}$ is the set of \textit{action state labels} $a^{t}_{u_j}$ for the ego user $u_i$’s neighbors $u_j$ at timestamp $t$, where $a^{t}_{u} \in \{0, 1\}$ indicates whether user $u$ has performed a social action (1) or not (0) by timestamp $t$. 
Additionally, the topology of the previously mentioned cascade sequence can be simplified by projecting it onto the time axis, denoted as $c(t_w)=\{(u_{j}, t_{j}) \}_{j\le w}$, where $(u_{j}, t_{j})$ indicates that user $u_{j}$ joined the cascade at timestamp $t_{j}$.
The diffusion network $G^D_i(t_w)=\{V^{P}_{i}, E^{D1}_{i}\}$ for the ego user $u_i$ is constructed from the subgraph of $G^S$ and is extended with the action state labels set $A^{t_w}_{u_i}$ at the observed timestamp $t_w$. The set $V^P_i=\{u_i, \Gamma^\rho_{u_i} \}$ includes the ego user $u_i$ and its $\rho$-neighbors $\Gamma^\rho_{u_i}$ along with their $A^{t}_{u_i}$.
In the next user prediction task, the diffusion network is formed by combining the interaction relations in $c(t_w)$ with the social topology in the social network $G^{S}$. 
Therefore, drawing from the symbol definitions provided above, next user prediction and social influence prediction can be defined as follows.
Therefore, based on the symbol definitions provided, next user prediction and social influence prediction can be defined as follows.
\begin{enumerate}[label=(\arabic*), leftmargin=0.3cm, itemsep=0cm, topsep=0cm, parsep=0cm, partopsep=0cm]
  \item[$\bullet$]\textit{Next user prediction}: Given the user set $U$, the social network $G^{S}=\{V^{U}, E^{S}\}$, the diffusion network $G^{D}(t_w)=\{V^{U}, E^{D1}\}$ of a post at observed time $t_w$ and a current cascade $c=\{(u_{1}, t_{1}), \\ \dots, (u_{p-1}, t_{p-1})\}$, next user prediction aims to predict the next spreading user $u_{p} \in U$ at predicted timestamp $t_p$.
  \item[$\bullet$]\textit{Social influence prediction}: Given the user set $U$ and the diffusion network $G^{D}_{i}(t_w)=\{V^{P}_{i}, E^{D1}_{i}\}$ of the ego user $u_i$ at the observed timestamp $t_w$, social influence prediction aims to predict the future action states $a^{t_p}_{u_i}$ of $u_i$ at predicted timestamp $t_p \gg t_w$.
\end{enumerate}

\subsubsection{Datasets analysis.}
\begin{itemize}[leftmargin=0.3cm, itemsep=0cm, topsep=0cm, parsep=0cm, partopsep=0cm]
    \item[$\bullet$] 
    In Table~\ref{tDS}, the source platforms contain social media, community Q\&A websites and new websites in next user prediction, and social media and academic citation websites in social influence prediction.  Community Q\&A websites offer context around user queries and responses, which can reveal user interests and influence future interactions. News websites, with their dynamic content and readership patterns, offer insights into how current events may drive user behavior.
    \item[$\bullet$] 
    The attributes of datasets for next user prediction are all characterized by the diffusion network, and the attributes for social influence prediction are all characterized by the social network and its derived diffusion network. These dual data capture both direct interactions and social influence patterns.
\end{itemize}

\subsubsection{Methodology.} We present a SOTA method for each subtask of microscopic information diffusion prediction, respectively. They briefly introduce their landmark algorithms.
\begin{itemize}[leftmargin=0.3cm, itemsep=0cm, topsep=0cm, parsep=0cm, partopsep=0cm]
    \item[$\bullet$]
    \textit{Next user prediction}: 
    \textbf{MCDAN}\cite{24MCDAN} predicts next user in cascades using a multi-scale context-enhanced dynamic attention network. It integrates global relationships from social networks and historical cascades, capturing user preferences with a multi-scale sequential hypergraph attention module. Next, a contextual attention enhancement module strengthens user interaction within cascades, while susceptibility labels are constructed based on user analysis. 
    Table~\ref{tb3.2.1} presents the performance comparison between MCDAN and baseline models on four datasets (Twitter-MSHGAT~\cite{sunMSHGAT2022}, Douban-ComSoc\cite{ComSoc2012}, Android~\cite{sunMSHGAT2022}, Christianity~\cite{sunMSHGAT2022}), measured by Hits@K and Map@K for K = 10, 50, 100.
    \item[$\bullet$]
    \textit{Social influence prediction}: 
    \textbf{FedInf}\cite{FedInf23} introduces a federated learning framework for social influence prediction, addressing privacy concerns and enabling cross-organizational collaboration. It uses differential privacy during model aggregation, projecting parameters into a lower-dimensional space to minimize noise. The whole framework consists of local training and global model updates. 
    Table~\ref{tb3.2.2} presents the performance comparison between FedInf and baseline models across three datasets (OAG-DeepInf~\cite{DeepInf18}, Digg-DeepInf~\cite{DeepInf18}, Higgs Twitter~\cite{de2013anatomy}), measured by AUC, precision, recall, and F1.
\end{itemize}

\section{Social bot detection}
\label{ch4}
The emergence of AI-driven \textit{social bots} that manipulate public opinion necessitates effective detection mechanisms to distinguish them from human accounts. Social bot detection is a binary classification problem where the bot label \(y^U \in \{0, 1\}\) indicates a bot user (1) or a human user (0). Current social bot detection tasks are categorized into three types based on input data: user-based, content-based, and graph-based detection~\cite{TwiBot22}.
\textbf{User-based bot detection} extracts features from user profiles and numerical or categorical data, employing traditional classifiers to identify bots~\cite{TwiBot22}. \textbf{Content-based bot detection} analyzes post content using NLP techniques, such as word embedding and RNN, on users' posts and descriptions~\cite{weiTwitter19}. \textbf{Graph-based bot detection} examines the user relationship structure within a social network \(G^{S}\), applying network analysis techniques to differentiate between bots and humans~\cite{TwiBot22}.

\subsection{Definition.}In light of the symbol definitions provided in the previous sections, user-based, content-based and graph-based bot detection can be defined as follows, respectively.
\begin{enumerate}[label=(\arabic*), leftmargin=0.3cm, itemsep=0cm, topsep=0cm, parsep=0cm, partopsep=0cm]
  \item[$\bullet$]\textit{User-based bot detection}: Given the labeled user set $U$ with user information, user-based bot detection aims to learn a classifier $f: f(U) \to y^U$ that can detect a user's bot label $y^U$.
  \item[$\bullet$]\textit{Content-based bot detection}: Given the labeled user set $U$ with user information and the post set $p$ with the post content, content-based bot detection aims to learn a classifier $f: f(U, P) \to y^U$ that can automatically detect the bot label $y^U$ for a user $u$.
  \item[$\bullet$]\textit{Graph-based bot detection}: Given the labeled user set $U$ with user information and the social network $G^{S}=\{V^{U}, E^{S}\}$, graph-based bot detection aims to learn a classifier $f: f(G^S)\to y^U$ that can automatically detect the bot label $y^U$ for a user $u$.
\end{enumerate}

\subsection{Datasets analysis.}
\begin{itemize}[leftmargin=0.3cm, itemsep=0cm, topsep=0cm, parsep=0cm, partopsep=0cm]
    \item[$\bullet$]
    In Table~\ref{tDS}, these datasets are widely used and all originated from Twitter platform. Twitter is a global social media platform that enables users to post and read short messages known as "tweets," allowing for the sharing of news, personal insights, and various types of content, while also facilitating interaction through following other users, replying, retweeting, and liking. 
    \item[$\bullet$]
    According to different input, each kind of social bot detection dataset captures different aspects of user information, propagation content, user behavior, and relations on Twitter.
\end{itemize}

\subsection{Methodology.} Similarly, we present a representative method for each subtask of social bot detection, respectively. They briefly introduce their landmark algorithms and current performance in this task.
\begin{itemize}[leftmargin=0.3cm, itemsep=0cm, topsep=0cm, parsep=0cm, partopsep=0cm]
    \item[$\bullet$]
    \textit{User-based and content-based bot detection}:
    \cite{KUDUGUNTA2018312} introduces a \textbf{Contextual LSTM network} for detecting social bots using user and content data. The model converts tweet text into GloVe vectors and combines them with user metadata. It employs synthetic minority oversampling and interprets LSTM hidden layers to differentiate tweets generated from human or bot. 
    Table~\ref{tb4.1-2} compares the performance of the Contextual LSTM and baseline models on account-level (user) and tweet-level bot detection tasks using the Cresci-2017 dataset~\cite{Cresci17}, measured by precision, recall, F1-score, accuracy, and AUC.
    \item[$\bullet$]
    \textit{Graph-based bot detection}:
    \cite{qxrBrother23} presents an \textbf{adversarial attack method} to bypass bot detection systems by placing a new bot near an existing one in the social graph. Utilizing a Relational Graph Convolutional Network (R-GCN), the method generates the new bot’s embedding and connects it to the target bot as second-order neighbors. An attribute recovery module conceals the new bot’s text attributes, achieving high attack success while differentiating the bot from human users. 
    Table~\ref{tb4.3} shows performance results on two datasets (Cresci-2015~\cite{CRESCI2015} and TwiBot-22~\cite{TwiBot22}), measured by attack success rate and new nodes detected as bots.
\end{itemize}

\section{Misinformation detection}
\label{ch5}
For the content during the information diffusion process, the decentralization of discourse has led to an increase in false or misleading information (i.e., \textit{misinformation}), which not only misguides the public but also threatens cyberspace security, making misinformation detection essential~\cite{QU2024102172, WANG2024102500}. Misinformation primarily includes rumors and fake news~\cite{VarshneyReview2021}. \textit{Rumors} are unverified, unintentionally spread on social media, while \textit{fake news} consists of deliberately false articles spread by official accounts or websites~\cite{BondielliSurvey2019}. Therefore, misinformation detection is divided into rumor and fake news detection.

Current methods rely on a single information source~\cite{RuchanskyCSI2017} by incorporating post content and contextual information. Specifically, \textit{social context} includes repost/comment content, user context (i.e., user information), and network context (i.e., user social network, diffusion network, and user engagement set)~\cite{BondielliSurvey2019, DouPreference21, MinDivide22, qiFakeSV2023, WangBots2024}.
Based on these, \textbf{rumor detection} determines the veracity label of a post \(y^P \in \{0, 1\}\), indicating a non-rumor ($0$) or a rumor ($1$). \textbf{Fake news detection} classifies news as real (0) or fake (1)~\cite{WuDECOR23}.

\subsection{Definition.}
Regarding rumor detection, the diffusion network $G^{D}_{i}=\{V^{P}_{i}, E^{D2}_{i}\}$ of each post comprises $V^{P}_{i}=\{p_{i}, r^i_{1}, r^i_{2}, \dots\} $ representing the set of a post $p_i\in P$ and its repost/comment $r^i_j$.
In graph-based fake news detection, diffusion network $G^{D}_{i}=\{V_{i}, E^{D3}_{i}\}$ for each news comprises $V_{i}=\{p_{i}, r^i_{1}, r^i_{2}, \dots\}$ including news $p_i$ and its engagement elements $r^i_j \in \{u^i_j, p^i_j, (u^i_j, p^i_j) \}$ (user, repost/comment, or both), varying across datasets.
Particularly, either user engagements set $M$ or diffusion network $G^{D}_{i}$ for each news can generate the \textit{news engagement graph} representing connections between news via overlapping engaged users.
User engagement sets in user information $M=\{m \in (u, p, k) | u\in U, p\in P \}$ indicates that user $u$ has $k$ engagements with news $p$ through reposts/comments~\cite{WuDECOR23}.
Therefore, in light of the symbol definitions provided, rumor detection and fake news detection can be defined as follows.
\begin{enumerate}[label=(\arabic*), leftmargin=0.3cm, itemsep=0cm, topsep=0cm, parsep=0cm, partopsep=0cm]
  \item[$\bullet$]\textit{Rumor detection}: Given the labeled post set $P$ with post content and several aspects of social context (the user set $U$ with user information, the social network $G^{S}=\{V^{U}, E^{S}\}$, the repost/comment content and the diffusion network $G^{D}_{i}=\{V^{P}_{i}, E^{D2}_{i}\}$ of each post), rumor detection aims to learn a classifier $f: f(p_i, U, G^S ,G^D_i) \to y^P$ that can automatically detect the veracity label $y^P$ of a post $p$.
  \item[$\bullet$]\textit{Fake news detection}: Given the labeled news set $P$ with the news content, the user set $U$ with the user information and several aspects of social context (the social network $G^{S}=\{V^{U}, E^{S}\}$, the repost/comment content, the diffusion network $G^{D}_{i}=\{V_{i}, E^{D3}_{i}\}$ of each news and the user engagements set $M$), fake news detection aims to learn a classifier $ f: f(P, U, G^S, G^D_i, M) \to y^P$ that can automatically detect the veracity label $y^P$ of a news $p$.
\end{enumerate}

\subsection{Datasets analysis.}
\begin{itemize}[leftmargin=0.3cm]
    \item[$\bullet$]
    In Table~\ref{tDS}, several datasets are processed from the same source data to enrich the data quality.
    Based on PHEME-v1, PHEME-v2 extends more events and PHEME-v3 offers the social network.
    Twitter15/16-RDMSC extra crawl the user information and friendships based on Twitter15/16.
    Compare to the FakeNewsNet, Fake-NewsNet-DECOR only utilizes news content and extra provides the raw user-news engagement records.
    FakeNewsNet-UPFD enhances data quality by incorporating news propagation networks and news retweet graphs and crawling near 20 million historical tweets of users.
    \item[$\bullet$]
    Regarding attributes, most rumor detection datasets include content. Three PHEME, Twitter15-RDMSC and Twitter16-RDMSC also contain user property. Datasets from FakeNewsNet to MC-Fake in the table have user engagement sets $M$ or the diffusion network $G^{D}_{i}$ for graph-based detection.
    \item[$\bullet$]
    Multimodal datasets for rumor detection are limited such as MR2. In contrast, fake news detection datasets are more diverse: FakeSV includes texts and videos, while MCFEND, Weibo21, Image-verification-corpus, Evons, FauxBuster, MM-Covid, and MuMIN combine texts and images, enhancing understanding and detection accuracy.
    \item[$\bullet$]
    Datasets use various labeling systems. Twitter15,Twitter16,Twitt-er15-RDMSC and Twitter16-RDMSC apply a four-class system: $y^P \in \{N, T, F, U\}$ (non-rumor, true rumor, false rumor, unverified rumor). MR2, Breaking!, and MuMIN use three categories: $y^P \in \{0, 1, 2\}$ (non-rumor, rumor, unverified rumor). FineFake and LIAR use six categories, while others employ a binary system: $y^P \in \{0, 1\}$ for rumor (non-rumor or rumor) and fake news (real or fake). These variations affect the granularity of detection models.
\end{itemize}

\subsection{Methodology.} The representative SOTA methods for the subtasks of misinformation detection are introduced in this paragraph respectively. They briefly introduce the landmark algorithms and current performance in this task.
\begin{itemize}[leftmargin=0.3cm, itemsep=0cm, topsep=0cm, parsep=0cm, partopsep=0cm]
    \item[$\bullet$]
    \textit{Rumor detection}:
    Graph-aware Multi-feature Interacting Network (GMIN)~\cite{YANG2024123687} detects rumors on social media by integrating text, user interactions, and propagation. It includes a Text-based Reasoning module that uses BERT and CNN-BiGRU for feature extraction, a Graph-aware Interaction module that constructs a user-text graph with GAT, a Propagation Structure module that applies GCN on diffusion graphs, and a Feature Collaboration module that fuses features via co-attention for early detection and interpretability.  
    Table~\ref{tb5.1} presents the performance comparison between GMIN and baseline models on three datasets (Ma-Weibo~\cite{ma2016detecting}, Twitter15~\cite{maDetect2017} and Twitter16~\cite{maDetect2017}), measured by precision, recall, F1-score, and accuracy.
    \item[$\bullet$]
    \textit{Fake news detection}:
    Adaptive Rationale Guidance (ARG) network~\cite{hu2024bad} employs Large and Small Language Models for fake news detection. It inputs news and rationales from an LLM, encodes them with BERT, and uses cross-attention to integrate rationales for classification, outperforming LLM-only and SLM-only methods.
    Table~\ref{tb5.2} presents the performance comparison between ARG and baseline models on two datasets (Weibo21~\cite{Weibo21} and GossipCop of FakeNewsNet~\cite{ShuFakeNewsNet2020}), measured by accuracy, F1-score and macro F1.
\end{itemize}

\section{Discussion}
\label{ch6}

\begin{figure}
    \centering
    \includegraphics[width=\linewidth]{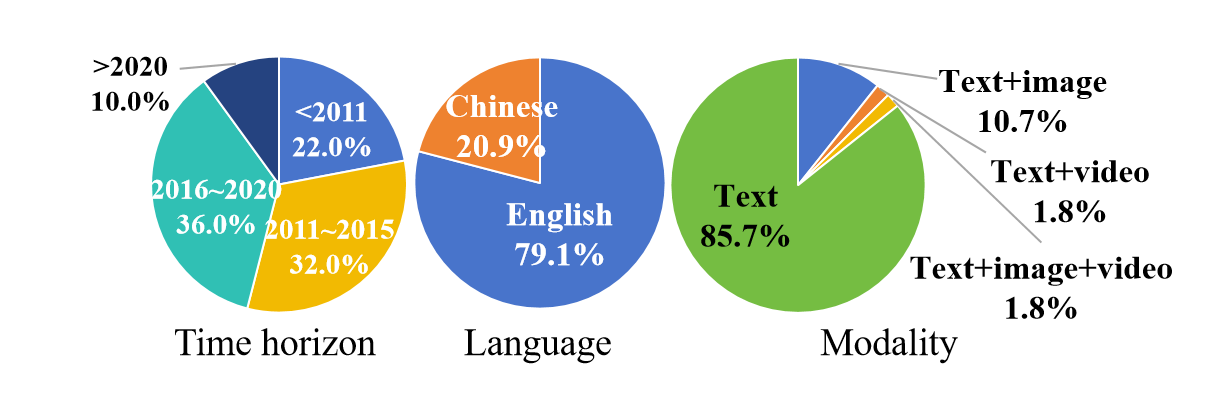}
    \caption{The statistics of time horizon, language and modality of all datasets.}
    \label{fg2}
\end{figure}

\subsection{Limitations on datasets} 
Currently, there is a lack of comprehensive datasets that encompass all six attributes, preventing simultaneous exploration of the three main tasks of information diffusion prediction. Furthermore, data analysis reveals that the time range covered by existing datasets is relatively outdated and that the data is limited in terms of language, modality, and source platform.

\begin{itemize}[leftmargin=0.3cm, itemsep=0cm, topsep=0cm, parsep=0cm, partopsep=0cm]
    \item[$\bullet$]
    Current datasets are incomplete due to the absence of user or content data. We identified six attributes for datasets in Section~\ref{2.2}.
    However, no dataset in Table~\ref{tDS} includes all six attributes although they often coexist in real scenarios. Therefore, creating datasets with comprehensive attributes is essential for accurately modeling propagation processes.

\item[$\bullet$]
Time-recorded datasets after 2020 are scarce, with those before 2016 comprising over half of the total and those from 2016 to 2020 making up the largest portion, as shown in Figure~\ref{fg2}. Information diffusion patterns from five to ten years ago may not apply today, making it crucial to acquire data from the last five years.

\item[$\bullet$]
This study acknowledges potential biases in various datasets. Nearly 80\% of current datasets focus on English-speaking platforms, leading to under-representation of non-English speakers, which can distort conclusions. Future work should prioritize collecting diverse datasets across different languages and regions.

\item[$\bullet$]
Most propagation content in the datasets is text, with only eight multimodal datasets available. Accurate results require diverse content types, such as images and videos.

\item[$\bullet$]
Most datasets are collected from a single platform, primarily Twitter. However, users engage across multiple platforms, and information diffusion patterns vary. Future research can collect data on the same users or events across multiple platforms for a more comprehensive understanding.

\item[$\bullet$]
Advances in generative AI allow for the creation of synthetic datasets simulating information diffusion processes. These technologies can generate users with varied attributes and set up different scenarios, including bot interference and AI-generated misinformation. These datasets will enable a deeper exploration of information diffusion complexities in the era of large language models.

\end{itemize}

\subsection{Limitations on research topics}  
Limitations and opportunities for future research topics are as follows.
\begin{itemize}[leftmargin=0.3cm, itemsep=0cm, topsep=0cm, parsep=0cm, partopsep=0cm]
    \item[$\bullet$]
    In addition to the prediction of popularity and user attitudes, the prediction of information diffusion is often based solely on cascade data, overlooking content-based information. Future work can incorporate the content, focusing on semantic and emotional attributes to enhance predictions of paths and effects.
    \item[$\bullet$]
    Current research on social bot detection focuses on identifying a single bot, while group bot detection methods is more crucial.
    \item[$\bullet$]
    Regarding misinformation detection, rumor detection generally overlooks early rumor detection for emerging events, and fake news detection lacks sufficient attention to the impact of temporal changes, user information, and social network, which can be focused more in the future.
    \item[$\bullet$]
    Existing studies~\cite{JiCommunity2023, BotMoE2023, WuSleep2024} typically study on a single platform or several independent platforms without links between different platforms, such as the same user, the same topic of content, etc. Future research is expected to consider cross-platform analysis to better explore the diffusion of the same user or event across different media.
\end{itemize}

\section{Conclusion}
\label{ch7}

In this survey, we categorize information diffusion tasks into three main categories based on the "5W Model" framework: information diffusion prediction, social bot detection, and misinformation detection. We review datasets within the information diffusion domain and assess them against six key attributes: user information, social network, bot labels, propagation content, diffusion network, and veracity labels. 
We further subdivide these main tasks into ten fine-grained subtasks, detailing their definitions, datasets analysis, and representative methods. Additionally, we highlight the limitations and future directions of current datasets and research topics in the field of information diffusion tasks. 
We anticipate that our survey will advance information diffusion research and provide valuable data support. Future research can address the limitations of our study by creating new datasets for information diffusion tasks and reviewing generative datasets produced by AI systems.


\bibliographystyle{ACM-Reference-Format}
\bibliography{Ref}

\clearpage
\appendix

\section{Performance of the methodology}
\label{chPerf}

To demonstrate the current experiment level, the performance in the corresponding landmark algorithms of all subtasks is shown in Table 3-11. The titles of tables present the datasets and the evaluation metrics used in those methods.

\section{Links of datasets}
\label{chAPP}

Datasets of general interest for information diffusion tasks are collected from the current works of each subtask in Table~\ref{tDS}, and can be openly found in the corresponding data repository with GitHub links or special organization links in Table~\ref{tURL} below, although several datasets need to apply for usage.
Their licenses can be found in the references and the URLs in this table.
With the purpose of privacy concerns and content security, the data have been processed during their creation stage, for example, by using the number instead of the user name.

\begin{table}[h]
    \caption{Performance comparison between CasFlow and baselines of cascade size prediction task on three datasets (Twitter-casflow, Sina Weibo, and APS) with different observation times. The best results in MSLE and MAPE are\textbf{bolded}.}
    \label{tb3.1.1}
    \centering
    \begin{tiny}
    
    \begin{tabular}{l|cccc|cccc|cccc} 
        \toprule
        \multirow{3}{*}{\textbf{Model}} & \multicolumn{4}{c|}{\textbf{Twitter-casflow}} & \multicolumn{4}{c|}{\textbf{Sina Weibo}} & \multicolumn{4}{c}{\textbf{APS}}  \\  \cmidrule(lr){2-5} \cmidrule(lr){6-9} \cmidrule(lr){10-13} 
        ~ & \multicolumn{2}{c}{\textbf{1 Day}} & \multicolumn{2}{c|}{\textbf{2 Days}} & \multicolumn{2}{c}{\textbf{0.5 Hour}} & \multicolumn{2}{c|}{\textbf{1 Hour}} & \multicolumn{2}{c}{\textbf{3 Years}} & \multicolumn{2}{c}{\textbf{5 Years}}   \\ \cmidrule(lr){2-3} \cmidrule(lr){4-5} \cmidrule(lr){6-7} \cmidrule(lr){8-9} \cmidrule(lr){10-11} \cmidrule(lr){12-13} 

        ~ & \tiny \textbf{MSLE} & \tiny \textbf{MAPE} & \tiny \textbf{MSLE} & \tiny \textbf{MAPE} & \tiny \textbf{MSLE} & \tiny \textbf{MAPE} & \tiny \textbf{MSLE} & \tiny \textbf{MAPE} & \tiny \textbf{MSLE} & \tiny \textbf{MAPE} & \tiny \textbf{MSLE} & \tiny \textbf{MAPE}   \\ 
        \midrule
        
        Feature-SH & 14.792 & 0.960 & 13.515 & 0.983 & 4.455 & 0.390 & 4.001 & 0.398 & 2.382 & 0.316 & 2.348 & 0.350   \\ 
        TimeSeries & 8.214 & 0.547 & 6.023 & 0.445 & 3.119 & 0.277 & 2.693 & 0.268 & 1.867 & 0.271 & 1.735 & 0.291   \\ 
        Feature-Linear & 9.326 & 0.520 & 6.758 & 0.459 & 2.959 & 0.258 & 2.640 & 0.271 & 1.852 & 0.272 & 1.728 & 0.291   \\ 
        Feature-Deep & 7.438 & 0.485 & 6.357 & 0.500 & 2.715 & 0.228 & 2.546 & 0.272 & 1.844 & 0.270 & 1.666 & 0.282   \\ 
        DeepHawkes & 7.216 & 0.587 & 5.788 & 0.536 & 2.891 & 0.268 & 2.796 & 0.282 & 1.573 & 0.271 & \textbf{1.324} & 0.335   \\ 
        CasCN & 7.183 & 0.547 & 5.561 & 0.525 & 2.804 & 0.254 & 2.732 & 0.273 & 1.562 & 0.268 & 1.421 & 0.265   \\ 
        DMT-LIC & 7.152 & 0.467 & 5.427 & 0.481 & 2.752 & 0.249 & 2.689 & 0.270 & 1.539 & 0.264 & 1.398 & 0.258   \\ 
        CasFlow* & \textbf{6.954} & \textbf{0.455} & \textbf{5.143} & \textbf{0.361} & \textbf{2.402} &\textbf{ 0.210} & \textbf{2.279} & \textbf{0.238} & \textbf{1.361} &\textbf{0.222} & 1.354* &\textbf{ 0.248}  \\ 
    \bottomrule
    \end{tabular}

    \begin{tablenotes}
    \item[1] A paired t-test is performed and * indicates a statistical significance $p < 0.001$ as compared to the best baselines.
    \end{tablenotes}
    \end{tiny}
\end{table}

\begin{table}[h]
    \caption{Performance comparison between RAGTrans and baselines of popularity prediction task on three datasets (SMPD, ICIP, and WeChat). The best results in MSE, MAE and SRC are \textbf{bolded}.}
    \label{tb3.1.2}
    \centering
    \begin{scriptsize}
    \begin{tabular}{l|ccc|ccc|ccc}
        \toprule
        \multirow{2}{*}{\textbf{Model}} & \multicolumn{3}{c|}{\textbf{SMPD}} & \multicolumn{3}{c|}{\textbf{ICIP}}  & \multicolumn{3}{c}{\textbf{WeChat}} \\ 
        \cmidrule(lr){2-4}  \cmidrule(lr){5-7} \cmidrule(lr){8-10} 
        ~ & \textbf{MSE} & \textbf{MAE} & \textbf{SRC} & \textbf{MSE} & \textbf{MAE} & \textbf{SRC} & \textbf{MSE} & \textbf{MAE} & \textbf{SRC} \\ 
        \midrule
        SVR & 4.9886 & 1.6749 & 0.5312 & 2.0942 & 1.0552 & 0.3723 & 2.9551 & 3.2072 & 0.0900 \\ 
        Hyfea & 4.9297 & 1.6623 & 0.5518 & 1.9813 & 0.9935 & 0.3641 & 2.8655 & 3.1073 & 0.1054 \\ 
        MFTM & 6.3697 & 1.9590 & 0.3479 & 1.6268 & 0.8923 & 0.4349 & 2.8104 & 3.0670 & 0.0794 \\ 
        DTCN & 4.2523 & 1.4998 & 0.5432 & 2.8361 & 1.3432 & 0.3893 & 3.6921 & 3.4432 & 0.0821 \\ 
        UHAN & 3.8471 & 1.4833 & 0.5541 & 2.7492 & 1.2824 & 0.3981 & 3.5925 & 3.3132 & 0.0835 \\ 
        MMVED & 6.3672 & 1.9607 & 0.2610 & 1.9831 & 1.0796 & 0.2606 & 2.9950 & 3.2151 & 0.0911 \\ 
        MGC & 5.5216 & 1.8489 & 0.3228 & 1.7706 & 1.0117 & 0.3906 & 2.945 & 3.1954 & 0.0891 \\ 
        MHF & 3.9297 & 1.5433 & 0.5419 & 1.8736 & 0.9132 & 0.4041 & 2.8351 & 3.0543 & 0.1019 \\ 
        CBAN & 5.6673 & 1.9058 & 0.1285 & 3.6143 & 1.3897 & 0.1294 & 2.9325 & 3.0945 & 0.0706 \\ 
        JAB & 6.1882 & 1.9359 & 0.2353 & 1.8606 & 0.9289 & 0.3057 & 2.9654 & 3.1185 & 0.0280 \\ 
        MASSL & 13.8925 & 3.1133 & 0.3037 & 1.8359 & 0.8809 & 0.3937 & 3.8951 & 3.1294 & 0.0529 \\ 
        HGNN & 5.1770 & 1.6061 & 0.4371 & 1.6711 & 0.9093 & 0.4423 & 2.9452 & 3.1753 & 0.0939 \\ 
        DHGNN & 5.0450 & 1.5836 & 0.4698 & 1.6493 & 0.9010 & 0.4556 & 2.9031 & 3.1048 & 0.0958 \\ 
        RAGTrans & \textbf{3.2763} & \textbf{1.3396} & \textbf{0.5859} & \textbf{1.2351} & \textbf{0.7149} & \textbf{0.5914} & \textbf{2.7928} & \textbf{2.9898} & \textbf{0.1147} \\ 

    \bottomrule
    \end{tabular}
    \end{scriptsize}
\end{table}

\begin{table}[h]
    \caption{Performance comparison between neutral state model and XYWZ1Z2 baseline model of user attitudes prediction task compared with the actual data on the COVID-19-rumor dataset. The best results in MAE and MSE are \textbf{bolded}.}
    \label{tb3.1.3}
    \centering
    \begin{scriptsize}
    \begin{tabular}{l|cccc|cccc}
    \toprule
        \multirow{2}{*}{\textbf{Model}} & \multicolumn{4}{c|}{\textbf{RMSE}} &  \multicolumn{4}{c}{MAE} \\ 
        \cmidrule(lr){2-9}
        ~ & $z0$ & $z1$ & $z2$& $sum$ & $z0$ &  $z1$ &  $z2$ &  $sum$ \\ 
        \midrule
        XYWZ1Z2 & 15504.88 & \textbf{279.15} & \textbf{133.07} & 15507.96 & 10130.19 &\textbf{193.52} &\textbf{96.43} &10420.14\\
        Neutral & \textbf{1284.82} & 387.77 & 253.01 & \textbf{1365.71} & \textbf{917.77} &266.27 &171.54& \textbf{1355.58}\\ 
    \bottomrule
    \end{tabular}
    \end{scriptsize}
\end{table}


\begin{table}[h]
    \caption{Performance comparison between MCDAN and baselines of next user prediction task on four datasets (Twitter-MSHGAT, Douban-ComSoc, Android, Christianity). The best results in Hits@K and Map@K for K = 10, 50, 100 are \textbf{bolded}.}
    \label{tb3.2.1}
    \centering
    \begin{scriptsize}
    \begin{tabular}{l|ccc|ccc|ccc|ccc}
    \toprule
        \multicolumn{13}{c}{\textbf{Hits@K}}\\ 
        \midrule
        \multirow{2}{*}{\centering \textbf{Model}} & \multicolumn{3}{c|}{\textbf{Twitter-MSHGAT}} &\multicolumn{3}{c|}{\textbf{Douban-ComSoc}}   & \multicolumn{3}{c|}{\textbf{Android}}& \multicolumn{3}{c}{\textbf{Christianity}} \\         \cmidrule(lr){2-4}  \cmidrule(lr){5-7} \cmidrule(lr){8-10} \cmidrule(lr){11-13} 
        ~ & @10 & @50 & @100 & @10 & @50 & @100 & @10 & @50 & @100 & @10 & @50 & @100 \\ 
        \midrule
        DeepDiffuse & 5.79 & 10.80 & 18.39 & 9.02 & 14.93 & 19.13 & 4.13 & 10.58 & 17.21 & 10.27 & 21.83 & 30.74  \\ 
        Topo-LSTM & 8.45 & 15.80 & 25.42 & 8.57 & 16.53 & 21.47 & 4.56 & 12.63 & 16.53 & 12.28 & 22.63 & 31.52  \\ 
        NDM & 15.21 & 28.23 & 32.30 & 10.00 & 21.13 & 30.14 & 4.85 & 14.24 & 18.97 & 15.41 & 31.36 & 45.86   \\ 
        SNIDSA & 25.37 & 36.64 & 42.89 & 16.23 & 27.24 & 35.59 & 5.63 & 15.22 & 20.93 & 17.74 & 34.58 & 48.76   \\ 
        FOREST & 28.67 & 42.07 & 49.75 & 19.50 & 32.03 & 39.08 & 9.68 & 17.73 & 24.08 & 24.85 & 42.01 & 51.28   \\ 
        Inf-VAE & 14.85 & 32.72 & 45.72 & 8.94 & 22.02 & 35.72 & 5.98 & 14.70 & 20.91 & 18.38 & 38.50 & 51.05   \\ 
        DyHGCN & 31.88 & 45.05 & 52.19 & 18.71 & 32.33 & 39.71 & 9.10 & 16.38 & 23.09 & 26.62 & 42.80 & 52.47   \\ 
        MS-HGAT & 33.50 & 49.59 & 58.91 & 21.33 & 35.25 & 42.75 & 10.41 & 20.31 & 27.55 & 28.80 & 47.14 & 55.62   \\ 
        Topic-HGAT & 35.12 & 51.41 & 61.15 & 23.50 & 37.58 & 45.66 & 11.76 & 21.72 & 29.39 & 30.02 & 48.73 & 57.80   \\ 
        RotDiff & 35.90 & 52.46 & 61.21 & 22.16 & 38.23 & 46.37 & 11.44 & 23.04 & 31.30 & 32.37 & 56.25 & 66.74   \\ 
        MCDAN & \textbf{38.45} & \textbf{55.78} & \textbf{64.25} & \textbf{49.39} & \textbf{58.58} & \textbf{62.81} & \textbf{11.89} & \textbf{25.10} & \textbf{32.79} & \textbf{35.49} & \textbf{56.92} & \textbf{67.41} \\ 
        
        \toprule
        \multicolumn{13}{c}{\textbf{Map@K}}\\ 
        \midrule
        \multirow{2}{*}{\centering \textbf{Model}}& \multicolumn{3}{c|}{\textbf{Twitter-MSHGAT}} &\multicolumn{3}{c|}{\textbf{Douban-ComSoc}}   & \multicolumn{3}{c|}{\textbf{Android}}& \multicolumn{3}{c}{\textbf{Christianity}} \\         \cmidrule(lr){2-4}  \cmidrule(lr){5-7} \cmidrule(lr){8-10} \cmidrule(lr){11-13} 
        ~ & @10 & @50 & @100 & @10 & @50 & @100 & @10 & @50 & @100 & @10 & @50 & @100 \\ 
        \midrule
        DeepDiffuse & 5.87 & 6.80 & 6.39 & 6.02 & 6.93 & 7.13 & 2.30 & 2.53 & 2.56 & 7.27 & 7.83 & 7.84   \\ 
        Topo-LSTM & 8.51 & 12.68 & 13.68 & 6.57 & 7.53 & 7.78 & 3.60 & 4.05 & 4.06 & 7.93 & 8.67 & 9.86   \\ 
        NDM & 12.41 & 13.23 & 14.30 & 8.24 & 8.73 & 9.14 & 2.01 & 2.22 & 2.93 & 7.41 & 7.68 & 7.86   \\ 
        SNIDSA & 15.34 & 16.64 & 16.89 & 10.02 & 11.24 & 11.59 & 2.98 & 3.24 & 3.97 & 8.69 & 8.94 & 9.72   \\ 
        FOREST & 19.60 & 20.21 & 21.75 & 11.26 & 11.84 & 11.94 & 5.83 & 6.17 & 6.26 & 14.64 & 15.45 & 15.58   \\ 
        Inf-VAE & 19.80 & 20.66 & 21.32 & 11.02 & 11.28 & 12.28 & 4.82 & 4.86 & 5.27 & 9.25 & 11.96 & 12.45   \\ 
        DyHGCN & 20.87 & 21.48 & 21.58 & 10.61 & 11.26 & 11.36 & 6.09 & 6.40 & 6.50 & 15.64 & 16.30 & 16.44   \\ 
        MS-HGAT & 22.49 & 23.17 & 23.30 & 11.72 & 12.52 & 12.60 & 6.39 & 6.87 & 6.96 & 17.44 & 18.27 & 18.40   \\ 
        Topic-HGAT & 23.71 & 24.53 & 24.66 & 12.70 & 13.61 & 13.72 & 6.80 & 7.53 & 7.68 & 18.98 & 19.85 & 19.99   \\ 
        RotDiff & 24.06 & 24.82 & 24.95 & 11.70 & 12.54 & 12.66 & 6.96 & 7.45 & 7.56 & 19.81 & 20.91 & 21.05   \\ 
        MCDAN & \textbf{25.89} & \textbf{26.69} & \textbf{26.81} & \textbf{40.70} & \textbf{41.13} & \textbf{41.19} & \textbf{7.47} & \textbf{8.04} & \textbf{8.15} & \textbf{22.88} & \textbf{23.78} & \textbf{23.94}  \\ 
    \bottomrule
    \end{tabular}
    
    \end{scriptsize}
\end{table}

\begin{table}[h]
    \caption{Performance comparison between FedInf and baselines of social influence prediction task across three datasets (OAG-DeepInf, Digg-DeepInf, Higgs Twitter). The best results in AUC, precision, recall and F1 are \textbf{bolded}.}
    \label{tb3.2.2}
    \centering
    \begin{scriptsize}
    \begin{tabular}{l|cccc|cccc|cccc}
    \toprule
        \multirow{2}{*}{\centering \textbf{Model}} & \multicolumn{4}{c|}{\textbf{OAG-DeepInf}}& \multicolumn{4}{c|}{\textbf{Digg-DeepInf}} & \multicolumn{4}{c}{\textbf{Higgs Twitter}} \\ 
        \cmidrule(lr){2-5} \cmidrule(lr){6-9} \cmidrule(lr){10-13} 
        ~ & \tiny \textbf{AUC} & \tiny \textbf{Pre} & \tiny \textbf{Recall} & \tiny \textbf{F1} & \tiny \textbf{AUC} & \tiny \textbf{Pre} & \tiny \textbf{Recall} & \tiny \textbf{F1} & \tiny \textbf{AUC} & \tiny \textbf{Pre} & \tiny \textbf{Recall} & \tiny \textbf{F1} \\ 
        \midrule
        DeepInf-GCN & 63.55 & 30.28 & \textbf{74.36} & 43.03 & 84.15 & 58.76 & 67.61 & 62.88 & 76.60 & 44.31 & 66.74 & 53.26 \\ 
        DeepInf-GAT & 72.84 & 41.18 & 63.02 & 49.81 & 90.13 & 66.82 & 74.46 & 70.44 & 79.68 & 48.12 & \textbf{68.09} & \textbf{56.93} \\ 
        HPPNP & 66.02 & 33.37 & 66.05 & 44.34 & 90.16 & \textbf{72.38} & 70.43 & 71.39 & 78.67 & 47.71 & 66.71 & 55.63 \\ 
        FedAvg* & 73.18 & \textbf{44.59} & 56.58 & \textbf{49.88} & \textbf{90.44} & 68.07 & \textbf{75.94} & \textbf{71.79} & \textbf{79.75} & \textbf{49.32} & 65.84 & 56.40 \\ 
        FedInf* & \textbf{73.19} & 43.51 & 58.22 & 49.81 & 90.26 & 70.08 & 69.89 & 69.99 & 79.73 & 48.85 & 66.02 & 56.15 \\ 
    \bottomrule
    \end{tabular}
    
    \begin{tablenotes}
    \item[1]  * indicates that the model operates in collaborative training mode; otherwise, it operates in \\centralized mode.
    \end{tablenotes}
    
    \end{scriptsize}
\end{table}


\newpage
\clearpage
\begin{table}[h]
    \caption{Performance comparison between Contextual LSTM and baselines of user-based and content-based bot detection bot detection tasks about the Cresci-2017 dataset. The best results in precision, recall, F1-score, accuracy and AUC are \textbf{bolded}}
    \label{tb4.1-2}
    \centering
    \begin{scriptsize}
    \begin{tabular}{l|l|ccccc} 

    \toprule
        \textbf{Task} & \textbf{Model} & \textbf{Pre} & \textbf{Recall} & \textbf{F1} & \textbf{Acc} & \textbf{AUC} \\ 
        \midrule
        \multirow{15}{*}{\rotatebox{90}{\makecell{User-based \\bot detection}}} & Logistic Regression & 0.94 & 0.93 & 0.93 & 0.91 & 0.89 \\ 
        ~ & SGD Classifier & 0.87 & 0.87 & 0.87 & 0.87 & 0.87 \\ 
        ~ & Random Forest Classifier & 0.98 & 0.98 & 0.98 & 0.98 & 0.98 \\ 
        ~ & AdaBoost Classifier & 0.98 & 0.98 & 0.98 & 0.98 & 0.98 \\ 
        ~ & 2-layer NN (500,200,1) RelU + Adam & 0.95 & 0.95 & 0.95 & 0.95 & 0.95 \\ 
        ~ & Logistic Regression (With SMOTENN) & 0.99 & 0.99 & 0.99 & 0.99 & 0.99 \\ 
        ~ & SGD Classifier (With SMOTENN) & 0.95 & 0.94 & 0.94 & 0.95 & 0.95 \\ 
        ~ & Random Forest Classifier (With SMOTENN) & 0.99 & 0.99 & 0.99 & 0.99 & 0.99 \\ 
        ~ & AdaBoost Classifier (With SMOTENN) & 1.00 & 1.00 & 1.00 & 1.00 & 1.00 \\ 
        ~ & 2-layer NN (300,200,1) RelU + Adam (With SMOTENN) & 0.99 & 0.99 & 0.99 & 0.99 & 0.98 \\ 
        ~ & Logistic Regression (With SMOTOMEK) & 0.92 & 0.91 & 0.91 & 0.91 & 0.91 \\ 
        ~ & SGD Classifier (With SMOTOMEK) & 0.90 & 0.90 & 0.90 & 0.90 & 0.90 \\ 
        ~ & Random Forest Classifier (With SMOTOMEK) & 0.99 & 0.99 & 0.99 & 0.99 & 0.99 \\ 
        ~ & AdaBoost Classifier (With SMOTOMEK) & 0.99 & 0.99 & 0.99 & 0.99 & 0.99 \\ 
        ~ & 2-layer NN (300,200,1) RelU+Adam (With SMOTOMEK) & 0.95 & 0.95 & 0.95 & 0.94 & 0.95 \\ 
        \midrule
        
        \multirow{17}{*}{\rotatebox{90}{\makecell{Content-based \\bot detection}}} & Logistic Regression (Metadata-only) & 0.80 & 0.80 & 0.79 & 0.80 & 0.76 \\ 
        ~ & SGD Classifier (Metadata-only) & 0.76 & 0.76 & 0.75 & 0.76 & 0.72 \\ 
        ~ & Random Forest Classifier (Metadata-only) & 0.80 & 0.80 & 0.80 & 0.80 & 0.78 \\ 
        ~ & AdaBoost Classifier (Metadata-only) & 0.80 & 0.80 & 0.79 & 0.80 & 0.76 \\ 
        ~ & Logistic Regression (Metadata-only+SMOTENN) & 0.92 & 0.92 & 0.92 & 0.92 & 0.88 \\ 
        ~ & SGD Classifier (Metadata-only+SMOTENN) & 0.91 & 0.90 & 0.90 & 0.90 & 0.89 \\ 
        ~ & Random Forest Classifier (Metadata-only+SMOTENN) & 0.92 & 0.92 & 0.92 & 0.92 & 0.88 \\ 
        ~ & AdaBoost Classifier (Metadata-only+SMOTENN) & 0.93 & 0.92 & 0.93 &0.92 & 0.91 \\ 
        ~ & Logistic Regression (Metadata-only+SMOTOMEK) & 0.79 & 0.77 & 0.76 & 0.77 & 0.77 \\ 
        ~ & SGD Classifier (Metadata-only+SMOTOMEK) & 0.78 & 0.77 & 0.76 & 0.77 & 0.77 \\ 
        ~ & Random Forest Classifier (Metadata-only+SMOTOMEK) & 0.79 & 0.77 & 0.77 & 0.77 & 0.77 \\ 
        ~ & AdaBoost Classifier (Metadata-only+SMOTOMEK) & 0.79 & 0.77 & 0.77 & 0.77 & 0.77 \\ 
        ~ & LSTM (Tweet-only+50D GloVE) & 0.96 & 0.96 & 0.96 & 0.96 & 0.96 \\ 
        ~ & Contextual LSTM (25D GloVE) & 0.96 & 0.96 & 0.96 & 0.96 & 0.96 \\ 
        ~ & Contextual LSTM (50D GloVE) & 0.96 & 0.96 & 0.96 & 0.96 & 0.96 \\ 
        ~ & Contextual LSTM(100D GloVE) & 0.96 & 0.96 & 0.96 & 0.96 & 0.96 \\ 
        ~ & Contextual LSTM (200D GloVE) & 0.96 & 0.96 & 0.96 & 0.96 & 0.96 \\ 
    \bottomrule
    \end{tabular}
    \end{scriptsize}
\end{table}

\begin{table}[h]
    \caption{Performance of the adversarial attack method in graph-based bot detection task on two datasets (Cresci-2015 and TwiBot-22). The best results in attack success rate and new node detected as bot are \textbf{bolded}.}
    \label{tb4.3}
    \centering
    \begin{footnotesize}
    \begin{tabular}{l|cc|cc} 
    \toprule
        \multirow{2}{*}{\centering \textbf{Model}} & \multicolumn{2}{c|}{\textbf{Cresci-2015}} & \multicolumn{2}{c}{\textbf{TwiBot-22}}  \\ 
        \cmidrule(lr){2-3} \cmidrule(lr){4-5} 
        ~ & \textbf{\makecell{Attack success \\rate}} & \textbf{\makecell{New node \\become bot}} & \textbf{\makecell{Attack success \\rate}} & \textbf{\makecell{New node \\become bot}}\\ 
        \midrule
        GCN & 95.68 ± 1.44 & 0.00 ± 0.00 & \textbf{93.97 ± 5.43} & 2.66 ± 5.09 \\ 
        HGT & 94.79 ± 1.18 & \textbf{0.06 ± 0.12} & 89.37 ± 3.56 & 5.40 ± 10.80 \\ 
        Simple-HGN & 95.74 ± 1.25 & 0.00 ± 0.00 & 74.94 ± 2.16 & 7.39 ± 14.78 \\ 
        R-GCN & \textbf{95.74 ± 1.50} & \textbf{0.06 ± 0.12} & 73.73 ± 1.71 & \textbf{12.94 ± 19.19} \\ 
    \bottomrule
    \end{tabular}
    \end{footnotesize}
\end{table}



\begin{table}[h]
    \caption{Performance comparison between GMIN and baselines of rumor detection task on three datasets (Ma-Weibo, Twitter15 and Twitter16). The best results in precision, recall, F1-score and accuracy are \textbf{bolded}.}
    \label{tb5.1}
    \centering
    \begin{footnotesize}
    \begin{tabular}{l|cccc|cc|cc} 

    \toprule
        \multirow{2}{*}{\centering \textbf{Model}} & \multicolumn{4}{c|}{\textbf{Ma-Weibo}} & \multicolumn{2}{c|}{\textbf{Twitter15}} & \multicolumn{2}{c}{\textbf{Twitter16}} \\ 
        \cmidrule(lr){2-5} \cmidrule(lr){6-7} \cmidrule(lr){8-9} 
        ~ & \textbf{F1} & \textbf{Rec} & \textbf{Pre} & \textbf{Acc} &\textbf{F1} & \textbf{Rec} & \textbf{Pre} & \textbf{Acc} \\ 
        \midrule
        Rumor2vec & 0.952 & 0.952 & 0.952 & 0.951 & 0.797 & 0.723 & 0.851 & 0.852 \\ 
        dEFEND & 0.913 & 0.915 & 0.913 & 0.914 & 0.654 & 0.738 & 0.631 & 0.702 \\ 
        HB-GAT & 0.955 & 0.954 & 0.954 & 0.955 & 0.919 & 0.920 & \textbf{0.951} & \textbf{0.951} \\ 
        BiGCN &\textbf{0.960}& 0.963 & 0.961 & 0.961 & 0.891 & 0.886 & 0.847 & 0.880 \\ 
        GCAN & 0.854 & 0.854 & 0.854 & 0.854 & 0.825 & 0.877 & 0.759 & 0.908 \\ 
        GLAN & 0.946 & 0.943 & 0.943 & 0.945 & 0.924 & 0.905 & 0.921 & 0.902 \\ 
        RvNN & 0.908 & 0.908 & 0.908 & 0.908 & 0.729 & 0.723 & 0.737 & 0.737 \\ 
        PPC & 0.920 & 0.926 & 0.923 & 0.921 & 0.811 & 0.842 & 0.820 & 0.863 \\ 
        GMIN & 0.959 & \textbf{0.963} & \textbf{0.957} & \textbf{0.961} & \textbf{0.931 }& \textbf{0.921} & 0.920 & 0.938 \\ 
    \bottomrule
    \end{tabular}
    \end{footnotesize}
\end{table}

\begin{table}[h]
    \caption{Performance comparison between ARG and baselines of fake news detection task on two datasets (Weibo21 and GossipCop of FakeNewsNet). The best results in accuracy, F1-score and macro F1 are \textbf{bolded}.}
    \label{tb5.2}
    \centering
    \begin{scriptsize}
    \begin{tabular}{ll|cccc|cccc} 

    \toprule
        \multicolumn{2}{c|}{\multirow{2}{*}{\centering \textbf{Method}} } & \multicolumn{4}{c|}{\textbf{Weibo21}} & \multicolumn{4}{c}{\textbf{GossipCop}} \\ 
        \cmidrule(lr){3-6} \cmidrule(lr){7-10} 
        ~ & ~ & \textbf{macF1} & \textbf{Acc} & \textbf{$F1_{real}$} & \textbf{$F1_{fake}$} & \textbf{macF1} & \textbf{Acc} & \textbf{$F1_{real}$} & \textbf{$Fl_{fake}$} \\ 
        \midrule
        \makecell{G1:LLM-\\Only} & GPT-3.5-turbo & 0.725 & 0.734 & 0.774 & 0.676 & 0.702 & 0.813 & 0.884 & 0.519 \\ 
        \midrule
        \multirow{4}{*}{\centering \makecell{G2: SLM-\\Only}} & Baseline & 0.753 & 0.754 & 0.769 & 0.737 & 0.765 & 0.862 & 0.916 & 0.615 \\ 
        ~ & EANNT & 0.754 & 0.756  & 0.773 & 0.736 & 0.763 & 0.864 & 0.918 & 0.608 \\ 
        ~ & Publisher-Emo & 0.761 & 0.763 & 0.784 & 0.738 & 0.766 & 0.868 & 0.920 & 0.611 \\ 
        ~ & ENDEF & 0.765 & 0.766  & 0.779 & 0.751 & 0.768 & 0.865 & 0.918 & 0.618 \\ 
        \midrule
        \multirow{4}{*}{\centering \makecell{G3: LLM\\+SLM}} & Baseline+Rationale & 0.767 & 0.769  & 0.787 & 0.748 & 0.777 & 0.870 & 0.921 & 0.633 \\ 
        ~ & SuperICL & 0.757 & 0.759  & 0.779 & 0.734 & 0.736 & 0.864 & 0.920 & 0.551 \\ 
        ~ & ARG &\textbf{0.784} & \textbf{0.786}  & \textbf{0.804} & \textbf{0.764} & \textbf{0.790} & \textbf{0.878} & \textbf{0.926} & \textbf{0.653} \\ 
        ~ & ARG-D & 0.771 & 0.772  & 0.785 & 0.756 & 0.778 & 0.870 & 0.921 & 0.634 \\ 
    \bottomrule
    \end{tabular}

    \begin{tablenotes}
    \item[1] ARG-D is the rationale-free ARG by distillation for cost-sensitive scenarios.
    \end{tablenotes}
    \end{scriptsize}
\end{table}

\setlength{\tabcolsep}{1.5pt}
\begin{table*}
\begin{threeparttable}
    \caption{The URLs of datasets for each subtask.}
    \label{tURL}
    \centering

    \begin{scriptsize}
    \begin{tabular} {m{0.7cm}|m{0.7cm}|m{3.2cm}|m{11cm}}

    \toprule 
    \multicolumn{2}{c|}{\textbf{Task}} & \textbf{Dataset} &  \textbf{URL} \\ 
    \midrule

    \multirow{14}{*}{\rotatebox{90}{\makecell[c]{Macroscopic information\\ diffusion prediction}}} &  \multirow{8}{*}{\rotatebox{90}{\makecell[c]{Cascade size\\ prediction}}} & Twitter-casflow~\cite{xuCasFlow2023}  & \url{https://github.com/Xovee/casflow} \\ 
    ~ & ~ & APS~\cite{xuCasFlow2023} & \url{https://github.com/Xovee/casflow}\\ 
    ~ & ~ & Sina Weibo~\cite{caoDeepHawkes2017} &\url{https://github.com/CaoQi92/DeepHawkes}\\ 
    ~ & ~ & Arxiv HEP-PH~\cite{Graphs05} & \url{http://snap.stanford.edu/data/cit-HepPh.html} \\ 
    ~ & ~ & Twitter-FOREST~\cite{YangMulti-Scale19} &\url{https://github.com/albertyang33/FOREST/tree/master/data}\\ 
    ~ & ~ & Douban~\cite{YangMulti-Scale19} & \url{https://github.com/albertyang33/FOREST/tree/master/data} \\ 
    ~ & ~ & Memetracker~\cite{YangMulti-Scale19} &\url{https://github.com/albertyang33/FOREST/tree/master/data}\\ \cmidrule(lr){2-4}
    
    ~ & \multirow{5}{*}{\rotatebox{90}{\makecell[c]{Popularity \\prediction}}} & SMPD~\cite{SMP19} & \url{https://smp-challenge.com/download.html}\\ 
    ~ & ~ & Yelp~\cite{linQuantify2022} &\url{https://www.yelp.com/dataset/} \\ 
    ~ & ~ & MovieLens~\cite{gargOnline2020} & \url{https://grouplens.org/datasets/movielens/}\\ 
    ~ & ~ & Micro-Videos~\cite{chenMicro2016} & \url{https://acmmm2016.wixsite.com/micro-videos/blank} \\ 
    ~ & ~ & MicroLens~\cite{niContent2023} & \url{https://github.com/westlake-repl/MicroLens}\\ 
    ~ & ~ & ICIP~\cite{ortis2019prediction} & \url{https://iplab.dmi.unict.it/popularitydataset/SIPD2020CHALLENGE/train/} \\
    \cmidrule(lr){2-4}
    
    ~ & \multirow{1}{*}{UAP\tnote{1}} & COVID-19-rumor~\cite{chengCOVID192021} & \url{https://github.com/MickeysClubhouse/COVID-19-rumor-dataset}\\ \midrule 
    
    \multirow{12}{*}{\rotatebox{90}{\makecell[c]{Microscopic information\\ diffusion prediction}}} & \multirow{6}{*}{\rotatebox{90}{\parbox{1.8cm}{\makecell[c]{Next user \\ prediction}}}} & Twitter-FOREST~\cite{YangMulti-Scale19} &\url{https://github.com/albertyang33/FOREST/tree/master/data} \\ 
    ~ & ~ & Douban-FOREST~\cite{YangMulti-Scale19} & \url{https://github.com/albertyang33/FOREST/tree/master/data} \\ 
    ~ & ~ & Memetracker~\cite{YangMulti-Scale19} &\url{https://github.com/albertyang33/FOREST/tree/master/data}\\ 
    ~ & ~ & Android~\cite{sunMSHGAT2022} &\url{https://github.com/slingling/MS-HGAT}\\ 
    ~ & ~ & Christianity~\cite{sunMSHGAT2022}  &\url{https://github.com/slingling/MS-HGAT} \\ 
    ~ & ~ & Twitter-MSHGAT~\cite{sunMSHGAT2022} & \url{https://github.com/slingling/MS-HGAT}\\ 
    ~ & ~ & Douban-MSHGAT~\cite{sunMSHGAT2022} & \url{https://github.com/slingling/MS-HGAT} \\ 
    ~ & ~ & Douban-ComSoc~\cite{ComSoc2012} & \url{http://www.cse.ust.hk/TL/dataset/Douban-50000.zip}\\ 
    \cmidrule(lr){2-4} 
    
    ~ &  \multirow{6}{*}{\rotatebox{90}{\parbox{1.3cm}{\centering \makecell[c]{Social influence\\ prediction}}}} & OAG-DeepInf~\cite{DeepInf18} &\url{https://github.com/xptree/DeepInf}\\ 
    ~ & ~ & Digg-DeepInf~\cite{DeepInf18} & \url{https://github.com/xptree/DeepInf} \\ 
    ~ & ~ & Twitter-DeepInf~\cite{DeepInf18} & \url{https://github.com/xptree/DeepInf} \\ 
    ~ & ~ & Higgs Twitter~\cite{de2013anatomy} & \url{https://snap.stanford.edu/data/higgs-twitter.html}\\ 
    ~ & ~ & Weibo-DeepInf~\cite{DeepInf18} &\url{https://github.com/xptree/DeepInf} \\ 
    ~ & ~ & Weibo-influencelocality~\cite{ZhangInfluence13}  & \url{http://www.aminer.cn/influencelocality}\\ \midrule 
    
     \multirow{8}{*}{\rotatebox{90}{\parbox{2.5cm}{\centering Social bot detection}}} &  \multirow{3}{*}{\rotatebox{90}{\parbox{0.8cm}{\centering \makecell[c]{User-\\based }}}} & Cresci-2017~\cite{Cresci17} & \url{https://botometer.osome.iu.edu/bot-repository/datasets.html}\\ 
    ~ & ~ & gilani-2017~\cite{Gilani17}  & \url{https://botometer.osome.iu.edu/bot-repository/datasets.html}\\ 
    ~ & ~ & botometer-feedback-2019~\cite{yangArming2019}  & \url{https://botometer.osome.iu.edu/bot-repository/datasets.html}\\ \cmidrule(lr){2-4} 
    
    ~ &  \multirow{2}{*}{\rotatebox{90}{\parbox{0.7cm}{\centering \makecell[c]{Content-\\based }}}} & PAN-AP-2019~\cite{rangel2019overview}  & \url{https://pan.webis.de/clef19/pan19-web/author-profiling.html}\\[0.1cm]
    ~ & ~ & caverlee-2011~\cite{leeSeven2021}  & \url{https://botometer.osome.iu.edu/bot-repository/datasets.html} \\[0.1cm] \cmidrule(lr){2-4} 
    
    ~ &  \multirow{3}{*}{\rotatebox{90}{\parbox{0.8cm}{\centering \makecell[c]{Graph-\\based }}}} & Cresci-2015~\cite{CRESCI2015}  & \url{https://botometer.osome.iu.edu/bot-repository/datasets.html} \\ 
    ~ & ~ & TwiBot-20~\cite{fengTwiBot20}  & \url{https://github.com/BunsenFeng/TwiBot-20}\\ 
    ~ & ~ & TwiBot-22~\cite{TwiBot22}  & \url{https://drive.google.com/drive/folders/1YwiOUwtl8pCd2GD97Q_WEzwEUtSPoxFs?usp=sharing}\\ \midrule 
    
     \multirow{23}{*}{\rotatebox{90}{Misinformation detection}} &  \multirow{9}{*}{\rotatebox{90}{Rumor detection}} & PHEME-v1~\cite{zubiagaLearning2016}  & \url{https://figshare.com/articles/dataset/PHEME_dataset_of_rumours_and_non-rumours/4010619?file=6453753} \\ 
    ~ & ~ & PHEME-v2~\cite{kochkinaAll2018}  & \url{https://figshare.com/articles/dataset/PHEME_dataset_for_Rumour_Detection_and_Veracity_Classification/6392078}  \\
    ~ & ~ & PHEME-v3~\cite{ZHENG2023}  & \url{https://www.sciencedirect.com/science/article/abs/pii/S0020025523006680}  \\ 
    ~ & ~ & Weibo-BiGCN~\cite{bian2020rumor}& \url{https://github.com/TianBian95/BiGCN}  \\ 
    ~ & ~ & Ma-Weibo~\cite{ma2016detecting}  & \url{https://www.dropbox.com/s/46r50ctrfa0ur1o/rumdect.zip?dl=0}  \\
    ~ & ~ & Twitter15~\cite{maDetect2017}  & \url{https://www.dropbox.com/s/7ewzdrbelpmrnxu/rumdetect2017.zip?dl=0} \\ 
    ~ & ~ & Twitter16~\cite{maDetect2017}  & \url{https://www.dropbox.com/s/7ewzdrbelpmrnxu/rumdetect2017.zip?dl=0} \\ 
    ~ & ~ & Twitter15-RDMSC~\cite{ZHENG2023}  & \url{https://github.com/Coder-HenryZa/RDMSC?tab=readme-ov-file}  \\ 
    ~ & ~ & Twitter16-RDMSC~\cite{ZHENG2023}  & \url{https://github.com/Coder-HenryZa/RDMSC?tab=readme-ov-file}  \\ 
    ~ & ~ & MR2~\cite{HuMR223} & \url{https://gitcode.com/THU-BPM/MR2/overview} \\
    \cmidrule(lr){2-4} 
    
    ~ &  \multirow{14}{*}{\rotatebox{90}{Fake news detection}} & FakeNewsNet~\cite{ShuFakeNewsNet2020}  & \url{https://github.com/KaiDMML/FakeNewsNet}\\ 
    ~ & ~ & FakeNewsNet-DECOR~\cite{WuDECOR23} & \url{https://github.com/jiayingwu19/DECOR} \\ 
    ~ & ~ & FakeNewsNet-UPFD~\cite{DouPreference21} & \url{https://github.com/safe-graph/GNN-FakeNews} \\ 
    ~ & ~ & TruthSeeker2023~\cite{Dadkhah23} & \url{https://www.unb.ca/cic/datasets/truthseeker-2023.html} \\ 
    ~ & ~ & MC-Fake~\cite{MinDivide22}  &\url{https://github.com/qwerfdsaplking/MC-Fake} \\ 
    ~ & ~ & FineFake~\cite{zhou2024finefake} & \url{https://github.com/Accuser907/FineFake} \\
    ~ & ~ & FauxBuster~\cite{FauxBuster18} & \url{https://ieeexplore.ieee.org/abstract/document/8622344} \\
    ~ & ~ & MM-Covid~\cite{li2020mmcovid} & \url{https://github.com/bigheiniu/X-COVID} \\
    ~ & ~ & MuMIN~\cite{MuMiN22} & \url{https://mumin-dataset.github.io/} \\

    ~ & ~ & CHECKED~\cite{yang2021checked} &\url{https://github.com/cyang03/checked} \\ 
    ~ & ~ & FakeSV~\cite{qiFakeSV2023} & \url{https://github.com/ICTMCG/FakeSV} \\ 
    ~ & ~ & FTT~\cite{huLearn2023} & \url {https://github.com/ICTMCG/FTT-ACL23} \\ 
    ~ & ~ & MCFEND~\cite{liMCFEND2024} & \url{https://github.com/Nicozwy/MCFEND_WWW/tree/main?tab=readme-ov-file} \\ 
    ~ & ~ & Weibo21~\cite{Weibo21} & \url{https://github.com/kennqiang/mdfend-weibo21} \\ 
    ~ & ~ & Image-verification-corpus~\cite{boididou2018detection} & \url{https://github.com/MKLab-ITI/image-verification-corpus} \\ 
    ~ & ~ & Breaking!~\cite{2019breaking} & \url{https://github.com/architapathak/FakeNewsCorpus} \\
    ~ & ~ & LIAR~\cite{LiarWang17} & \url{https://www.cs.ucsb.edu/~william/data/liar_dataset.zip} \\
    ~ & ~ & Evons~\cite{2022evons} & \url{https://github.com/krstovski/evons} \\
     ~ & ~ & WeChat~\cite{WeakWang2020} & \url{https://github.com/yaqingwang/WeFEND-AAAI20} \\
     ~ & ~ & Fakeddit~\cite{nakamura2020fakeddit} & \url{https://github.com/entitize/Fakeddit} \\
    
    \bottomrule
    \end{tabular}
    
    \begin{tablenotes}
    \item[1] UAP: User Attitudes Prediction
    \end{tablenotes}
    
\end{scriptsize}
\end{threeparttable}
\end{table*}

\end{document}